\documentclass[sigconf]{acmart}

\usepackage{booktabs} 
\usepackage[margin=4pt,caption=false]{subfig}
\usepackage{diagbox}
\usepackage{multirow}
\usepackage{algorithm}
\usepackage{graphicx}
\usepackage{amsmath}
\usepackage{algpseudocode}
\usepackage{tikz}
\newcommand*\circled[1]{\tikz[baseline=(char.base)]{
            \node[shape=circle,draw,inner sep=0pt] (char) {#1};}}
\DeclareRobustCommand\circled[1]{\tikz[baseline=(char.base)]{
            \node[shape=circle,draw,inner sep=0pt] (char) {#1};}}

\algblockdefx[Foreach]{Foreach}{EndForeach}[1]{\textbf{foreach} #1 \textbf{do}}{\textbf{end foreach}}
\algnewcommand\algorithmiccase{\textbf{case}}
\algdef{SE}[CASE]{Case}{EndCase}[1]{\algorithmiccase\ #1}{\algorithmicend\ \algorithmiccase}%
\algtext*{EndCase}%
\makeatletter
\def\algbackskip{\hskip-\ALG@thistlm}
\makeatother
\newenvironment{myproof}[1][\proofname]{\proof[#1]\mbox{}}{\endproof}
\usepackage[titletoc,title]{appendix}

\copyrightyear{2017} 
\acmYear{2017} 
\setcopyright{acmcopyright}
\acmConference{CCS'17}{}{Oct. 30--Nov. 3, 2017, Dallas, TX, USA.}
\acmPrice{15.00}
\acmDOI{http://dx.doi.org/10.1145/3133956.3134019}
\acmISBN{ISBN 978-1-4503-4946-8/17/10} 

\begin{document}
\title{Be Selfish and Avoid Dilemmas: 
\\Fork After Withholding (FAW) Attacks on Bitcoin}

\author{Yujin Kwon}
\affiliation{\institution{KAIST}}
\email{dbwls8724@kaist.ac.kr}

\author{Dohyun Kim}
\affiliation{\institution{KAIST}}
\email{dohyunjk@kaist.ac.kr}

\author{Yunmok Son}
\affiliation{\institution{KAIST}}
\email{yunmok00@kaist.ac.kr}

\author{Eugene Vasserman}
\affiliation{\institution{Kansas State University}}
\email{eyv@ksu.edu}

\author{Yongdae Kim}
\affiliation{\institution{KAIST}}
\email{yongdaek@kaist.ac.kr}

\renewcommand{\shortauthors}{K. Yujin et al.}

\begin{abstract}
In the Bitcoin system, participants are rewarded for solving cryptographic puzzles.
In order to receive more consistent rewards over time, some
participants organize mining pools and split the rewards from the pool in proportion to each participant's contribution.
However, several attacks threaten the ability to participate in pools. The \textit{block withholding} (BWH) attack makes the pool reward system unfair by letting
malicious participants receive unearned wages while only pretending to contribute work.
When two pools launch BWH attacks against each other, they encounter the
\textit{miner's dilemma}: in a Nash equilibrium, the revenue of both pools is diminished.
In another attack called \textit{selfish mining}, an attacker can unfairly
earn extra rewards by deliberately generating forks.

In this paper, we propose a novel attack called a \textit{fork after withholding} (FAW) attack.
FAW is not just another attack. The reward for an FAW attacker
\emph{is always equal to or greater than that for a BWH attacker}, and it is 
usable up to four times more often per pool than in BWH attack.
When considering multiple pools --- the current state of the Bitcoin network --
the extra reward for an FAW attack is about 56\% more than that for a BWH attack.
Furthermore, when two pools execute FAW attacks on each other, the miner's dilemma
may not hold: under certain circumstances, the larger pool can consistently win.
More importantly, an FAW attack, while using intentional forks, does not suffer from practicality issues, unlike selfish mining.
We also discuss partial countermeasures against the FAW attack,
but finding a cheap and efficient countermeasure remains an open problem.
As a result, we expect to see FAW attacks among mining pools.
\end{abstract}

%
%



\maketitle

\section{Introduction}
\label{Introduction}

Bitcoin is the first fully decentralized cryptocurrency~\cite{nakamoto2008bitcoin}.
Its value has increased significantly as has its rate of adoption since its deployment in 2009~\cite{value}.
The security properties of Bitcoin rely on \textit{blockchain}
technology~\cite{blockchain},
which is an open ledger containing all current and historical transactions in the system.
To prevent alterations of previous transactions and maintain 
the integrity of the ledger, the system requires
participants to use their computational power to generate \textit{proofs of work} (PoWs) by solving cryptographic puzzles.
A PoW is required to generate a block and add transactions to the blockchain.
After someone generates a block by solving the puzzle, and this solution
is propagated to the Bitcoin network,
a new round starts and all nodes begin solving a new cryptographic puzzle.
The process of block generation is called \textit{mining}, and those carrying out this activity are called \textit{miners}.

As of May 2017, a miner who solves a puzzle
is rewarded with 12.5 bitcoins (BTC).
The average time for each round (time to solve the puzzle) is intended to be constant (10 minutes), 
so mining
difficulty is adjusted automatically about every two weeks.
As mining difficulty increases,
solo miners may have to wait for a long time, on average,
to receive any reward.
To prevent this reward ``starvation,'' some miners have organized \textit{mining pools} that engage in profit sharing.
Most pools consist of a \textit{pool manager} and \textit{worker} miners.
The manager runs the Bitcoin protocol, acting as a single node, but miners join a pool
by connecting to the pool's protocol~\cite{stratum_wiki} instead of directly joining Bitcoin.
A pool manager forwards unsolved work units to miners, who then generate partial proofs of work (PPoWs) and full proofs of work (FPoWs), and submit them to the manager as \textit{shares}.
If a miner generates an FPoW and submits it to the manager, the manager
broadcasts a block generated from the FPoW to the Bitcoin system, receives the
reward, and distributes the reward to participating miners.
Each miner is paid based on the fraction of shares contributed 
relative to the other miners in the pool. 
Thus, participants are rewarded based on PPoWs, which
have absolutely no value in the Bitcoin system.
The Bitcoin network currently consists of solo miners, open pools that allow any miner to join,
and closed (private) pools that require a private relationship to join.

There are several attacks on Bitcoin~\cite{karame2012double,
eyal2014majority, rosenfeld2011analysis}; our work focuses on
two well-known mining attacks: \textit{selfish mining} and
\textit{block withholding}.
Selfish mining abuses Bitcoin's \textit{forks} mechanism to derive an unfair reward.
A fork can occur when at least two cryptographic solutions (blocks) are propagated in a round. This may occur when solutions are discovered almost simultaneously, and take time to propagate through the Bitcoin network.
Only one branch of a fork can be valid (only one solution will be accepted);
others are eventually invalidated.
In selfish mining, proposed by Eyal et al. in 2014~\cite{eyal2014majority}, 
an attacker does not propagate a block immediately,
but generates forks intentionally by propagating a block selectively
only when another honest miner generates a block.
The attacker can earn a greater reward by invalidating honest miners' blocks
if she has enough computational power.

In a Block Withholding (BWH) attack, a miner in a pool submits only PPoWs, but not FPoWs. When an attacker launches a BWH attack against a single pool
and conducts honest mining with the rest of her computational power, she earns an extra reward, while the target pool takes a loss.
All pools are still vulnerable to this attack because no
efficient and cheap defense has emerged, despite ongoing research.
In 2015, Eyal~\cite{eyal2015miner} first modeled a game between two
BWH attacking pools, and discovered the \textit{miner's dilemma}: when two pools
attack each other, both will take a loss in equilibrium.
This is analogous to the classic ``prisoners' dilemma''.
Currently, pools implicitly agree not to launch BWH attacks against each other
because it would harm everyone. 
In other words, while BWH attack is
always profitable, the BWH attack game is not.
We describe these two attacks in more detail in Section~\ref{Preliminaries}.

In this paper, we describe a new attack called a \textit{fork after
withholding} (FAW) attack, which combines a BWH attack 
with intentional forks.
Like the BWH attack, the FAW attack is always profitable regardless of an
attacker's computational power or network connection state.
The FAW attack also provides superior rewards compared to the BWH attack -- in fact, the BWH attacker's reward is the \emph{lower bound} of the FAW attacker's.
We analyze both the single- and multi-pool FAW attack variants in Sections~\ref{Onepool} and~\ref{multipool}, respectively.
Then, in Section~\ref{Game}, we model the \textit{FAW attack game} between two FAW attacking pools
and discover that the attack becomes a \textit{size game between the two pools}, breaking the miner's
dilemma stalemate.

\vspace*{2mm}
\noindent\textbf{Single-pool FAW attack.} Like the BWH attacker, an FAW attacker joins the target pool and 
executes an FAW attack against it. The node submits FPoWs to
the pool manager \textit{only when another miner, neither the attacker
nor a miner in the target pool, generates a block}. If the pool manager
accepts the submitted FPoW,
he propagates it, and a fork will be generated.
Then, all Bitcoin network participants must select one branch.
If the attacker's block is selected, the target pool receives the reward,
and she is also rewarded from the pool.
When attacking a single pool, an FAW attacker can earn
extra rewards in any case. The lower bound of the extra reward is
that for a BWH attacker. In Section~\ref{Onepool}, we show quantitatively that the FAW
attacker can earn extra rewards one to four times more than that for the
BWH attacker in a large pool (representing 20\% of the computational power
of the entire Bitcoin network).

\vspace*{2mm}
\noindent\textbf{Multi-pool FAW attack.}
To increase her reward, she can simultaneously attack multiple
pools, so we expand our
attack to consider the FAW attack against $n$
pools. As in the single pool case, our analysis shows that the FAW attack is
always profitable, and that the FAW attacker earns a greater 
reward than the BWH attacker.
If an attacker executes the FAW attack against four pools that are currently popular~\cite{hashdist}, her extra reward will be about 56\% greater than that for
the BWH attacker. Note that the extra reward for attacking multiple pools
is more than that for a single pool attack.
Details of the multi-pool attack analysis are presented in Section~\ref{multipool}.

\vspace*{2mm}
\noindent\textbf{FAW attack game.}
Section~\ref{Game} considers a scenario in which two pools execute
FAW attacks against each other.
There is a Nash equilibrium in the game;
however, unlike in the BWH attack game~\cite{eyal2015miner}, there exists a 
condition in which the larger pool always earns the extra reward. 
That is, the miner's dilemma may not hold.
Therefore, the equilibrium for the FAW attack game in which two pools decide
whether to attack may be a Pareto optimal.

\vspace*{2mm}
\noindent\textbf{FAW attack vs. selfish mining.}
We also compare the FAW attack to selfish mining~\cite{eyal2014majority} in Section~\ref{Compare}.
Selfish mining is not always profitable, and the attacker is easily detectable.
Moreover, selfish mining is known to be impractical~\cite{impractical,gervais2016security,bonneau2015sok}.
Indeed, previous attacks on mining that generate intentional forks 
share these properties, making them impractical. 
However, unlike selfish mining, the FAW
attack is always profitable, and detecting FAW attackers is harder than
detecting selfish mining attackers even though the
FAW attack does utilize intentional forks. 

\vspace*{2mm}
In Section~\ref{Practicality}, we discuss various parameters used
throughout the study, some of which can be computed in
advance, making FAW attacks feasible. 
One specific parameter is hard to
compute in advance, but we show that the FAW attack is
still profitable even without knowing it.
Moreover, it is possible to identify Sybil nodes, but not the attacker.
Though we also propose several possible countermeasures, including
a method for detecting FAW attacks in Section~\ref{Discussion},
\textit{we find no practical defense} for FAW attacks. 

\vspace*{2mm}
\noindent\textbf{Contributions.}
This paper makes the following contributions:
\begin{enumerate}
\item We propose the FAW attack, which is always
profitable (unlike selfish mining) regardless of the attacker's computational power and network capability.
The extra reward for an FAW attack is always at least as high as that for a BWH attack.
\item We analyze the FAW attack when the attack target is one pool and
generalize to an attack against $n$ pools. Moreover, we consider an
FAW attack pool game, in which two pools execute FAW attacks
against each other. We prove that it can give rise to a
pool size game, deviating from the miner's dilemma that exists in the BWH attack.
\item We discuss and propose partial countermeasures for 
preventing an FAW attack. However, these
defenses are neither perfect nor practical, leaving an open problem.
\end{enumerate}

\section{Preliminaries}
\label{Preliminaries}

Although built with security in mind, Bitcoin is vulnerable to several 
attacks that allow an attacker to unfairly earn additional profits at others'
expense.
In this section, we describe Bitcoin and
the existing attacks against it that are related to our attack.

\subsection{Bitcoin Basics}

\noindent\textbf{Mining Process: }
The header of each block in a blockchain contains a Merkle
root~\cite{merkle1980protocols} of the latest transactions, the hash value of the
previous block header, and a nonce. In the Bitcoin system, ``mining'' is the process of
generating nonces, which are PoWs derived from solving cryptographic
puzzles. This work is performed by peers, known as ``miners''.
In short, a miner must find a valid nonce as a PoW satisfying 
$sha256(sha256(blkhdr))<t$,
where $blkhdr$ refers to all data in a block header, and
$t$ is a 256-bit number specified by the Bitcoin protocol, so 
it is more difficult to find a valid nonce given a smaller $t$.
The value of $t$ is automatically adjusted by the Bitcoin system 
to keep the average duration of each
round 10 minutes.
When a miner finds a valid nonce and generates a new block, this block is
broadcast to every node in the Bitcoin network. When another node receives it,
the node regards this block as the new head of the blockchain. 
At the time of writing,
a miner receives 12.5~BTC
as a reward for solving the puzzle
and extending the blockchain
at the expense of computational power.

\smallskip\noindent\textbf{Forks: }
If two miners independently build and broadcast two different valid
blocks,
a node may consider the block first received as the new blockchain head.
Because of different network
latencies~\cite{decker2013information}, more than two heads
can exist at the same time. This situation is called
a \textit{fork}. 
By appending a subsequent block to only one branch in the fork, 
the branch is defined as valid, while all others are invalidated.
Moreover, forks can also be intentionally generated.
When an attacker generates a block, she can withhold it
until another miner generates and propagates another block. Then,
the attacker can propagate her block right after she
listens to the block propagation, intentionally causing a fork, for
double-spending~\cite{doublespend} or selfish mining~\cite{eyal2014majority,
nayak2016stubborn, sapirshtein2015optimal} attacks.


\smallskip
\noindent\textbf{Mining Pools: }
Because successfully generating blocks requires a non-trivial amount of luck,
mining pools have been organized to reduce variance in the miners'
rewards as mining difficulty increases. Most mining pools consist of a manager and multiple miners.
At the start of every round, the manager distributes work to the miners~\cite{stratum_wiki}, and every miner uses his computing power to generate either partial (PPoW) or full (FPoW) PoWs.
The difficulty of generating a PPoW is lower than that of an FPoW.
For example, the hash value of a block header can have a 32-bit and 
72-bit zero prefix in a PPoW and in an FPoW, respectively.
When a miner generates a PPoW or an FPoW, he submits it as a share.
If a miner is lucky enough to generate an FPoW, the manager propagates it
and receives a reward, which he shares with
the miners in proportion to their submissions.

\subsection{Related Work}
\label{Related}
We review two related attacks on Bitcoin mining and new Bitcoin
protocol designs in this section. 

\vspace*{2mm}
\noindent\textbf{Selfish Mining: } 
Selfish mining~\cite{eyal2014majority, bahack2013theoretical}
generates forks intentionally.
If an attacker generates an FPoW, she does not propagate it immediately.
As soon as another miner propagates a block, the attacker selectively
propagates her withheld blocks according to their number 
to generate a fork.
This fork may invalidate honest miners' blocks, and the
attacker can improperly earn an extra reward. However, because the
attacker can also lose her block if her branch is not chosen, she must have 
greater computational power to make selfish mining
profitable, especially if her network connection capability is low~\cite{eyal2014majority}.
Many researchers have investigated selfish mining.
Sapirshtein et al.~\cite{sapirshtein2015optimal} and Nayak et al.~\cite{nayak2016stubborn} showed that
the original selfish mining scheme is not optimal and provided a new algorithm to optimize the selfish mining.
The former study~\cite{sapirshtein2015optimal} modeled an optimal selfish mining strategy using
the delay parameter of the Bitcoin network rather than the attacker's network capability.
It also stated that a profitable selfish miner can execute a double spending attack.
Nayak et al.~\cite{nayak2016stubborn} extended the parameters used for 
selfish mining strategy
and combined selfish mining with a network-level eclipse attack.
Although powerful, selfish mining is widely considered to be impractical~\cite{impractical, nayak2016stubborn}.
 Carlsten et al. studied selfish mining under a transaction fee regime 
(a Bitcoin reward system for the far future) and improved the attack by considering 
a variable reward for each block~\cite{carlsten2016instability}.
Selfish mining and FAW attacks are compared in Section~\ref{Compare}.

\vspace*{2mm}
\noindent\textbf{BWH Attack: }
The BWH attack was introduced by
Rosenfeld~\cite{rosenfeld2011analysis}. 
An attacker joins a target pool and then 
submits only PPoWs, but not FPoWs, unlike 
honest pool miners.
Because the attacker pretends to contribute to the target pool and gets 
paid, the pool suffers a loss.
Courtois et
al.~\cite{courtois2014subversive} generalized the concept of the BWH attack,
considering an attacker who mines both solo and in
pools. They showed that the attacker can unfairly earn a greater reward
through a BWH attack. 
This attack was carried out against the
``Eligius'' mining pool in 2014, with the pool losing
300 BTC~\cite{eligius}. In this case, the
manager found the attacker, who was using only two
Bitcoin accounts and did not submit FPoWs for an extended period of time.
If the attacker had used many more Bitcoin accounts, distributing
computational power across them and
masquerading as many workers, each of whom would mine 
in the pool for only a short time before being replaced with a new account, 
the manager may not have
detected her. Meanwhile, managers can always notice whether a BWH
attack has occurred by comparing the number of submitted PPoWs and FPoWs. 
However, managers \textit{cannot prevent the
attack}.
In 2015, Luu et
al.~\cite{luu2015power} found the optimal BWH attack strategy against one pool and multiple pools by defining the
power splitting game. Eyal~\cite{eyal2015miner} modeled
the BWH attack game between two mining pools. This study showed that such a game results in the miner's dilemma, which is analogous to the prisoner's
dilemma, because it creates mutual loss in the Nash equilibrium.
We propose the FAW attack, which improves the BWH attack. 
The FAW attack gives an attacker extra rewards up to four times more than 
those for a BWH attacker.
Moreover, we show that the miner's dilemma may not hold in the FAW attack game.
FAW and BWH attacks can occur against Ethereum~\cite{wood2014ethereum}, Litecoin~\cite{litecoin}, Dogecoin~\cite{dogecoin}, and Permacoin~\cite{miller2014permacoin} as well as Bitcoin.

\vspace*{2mm}
\noindent\textbf{New Bitcoin Protocols: } 
Many papers have proposed new protocols to solve various problems with Bitcoin such as selfish mining,
double spending, and scalability~\cite{sompolinsky2015secure,eyal2016bitcoin,luu2016secure,kogias2016enhancing}.
To prevent BWH attacks, 
several works~\cite{rosenfeld2011analysis, twophase} have proposed new 
two-phase PoW protocols, 
dividing work into two smaller cryptographic puzzles.
Then, a manager gives one puzzle to miners in his pool and 
solves the other himself.
As a result, miners cannot know whether their solutions are FPoWs 
and cannot execute BWH attacks. 
Under these protocols, an FAW attack also cannot happen.
However, Bitcoin participants do not want to adopt them, for reasons
described in Section~\ref{Countermeasure}.
Luu et al. ~\cite{luusmart} proposed a decentralized pool protocol called  \textit{SmartPool} 
that applies \textit{smart contracts}.  
They argued that attacks on pools would no longer be profitable 
if SmartPool exists as only one mining pool in the Bitcoin system. 
However, SmartPool's full adoption is considered to be a long way off~\cite{smart}.
We discuss other possible defense mechanisms against an 
FAW attack in Section~\ref{Countermeasure}.
\vspace*{-3mm}

\section{Attack Model and Assumptions}
\label{Model}

In this section, we specify our attack model and the assumptions made in
the rest of the paper.

\subsection{Attack Model}

First, an attacker can be a solo miner, or the manager of a closed or
open mining pool. Second, the attacker can launch Sybil attacks~\cite{babaioff2012bitcoin}, i.e.,
the attacker can generate an arbitrary number of identities and join multiple open pools with different IDs and Bitcoin accounts.
However, we assume that the attacker cannot join closed pools since those
require private information.
Third, the computational power of an attacker is finite, and she can distribute it into any fraction for both \textit{innocent
mining} (i.e., working as an honest solo miner) and \textit{infiltration mining} 
(i.e., joining and mining in multiple open pools to gain extra illicit rewards).
If an attacker is the manager of an open pool, her infiltration mining power
(the computational power used for infiltration mining)
should be loyal mining power~\footnote{Defined by Eyal
as ``mining power \ldots either run directly by the pool owners or sold as a service but run
on the pool owners' hardware''~\cite{eyal2015miner}.}
(the amount of loyal mining power pools possess is generally a trade
secret~\cite{eyal2015miner}).
Finally, the rushing adversary can
plant many Sybil nodes in the Bitcoin network, which can simply listen to the propagation of valid blocks and propagate the attacker's block preferentially when the attacker's block and another block are released simultaneously.
By this means, the attacker can track the propagation of other blocks and propagate her own as fast as possible using Sybil nodes.
Note that these nodes require very little computational power because their role is only to listen and propagate a block;
thus planting Sybil nodes involves negligible computation cost for the
attacker.

\subsection{Assumptions}
For the sake of simplicity, we make the following assumptions, consistent with other
attacks on Bitcoin mining~\cite{eyal2015miner, eyal2014majority, luu2015power}:

\begin{enumerate}
\item The normalized total computational power of the Bitcoin system is 1.
Therefore, any computational power is represented as a fraction of this total. Also, we assume that the computational power of any one
miner or mining pool is less than 0.5 to prevent a ``51\% attack'' on the
Bitcoin network~\cite{bradbury2013problem}. 

\item No managers or miners, except FAW attackers, launch attacks. 
We do not consider other attacks, such as BWH attacks or selfish mining,
alongside the FAW attack. 

\item  The reward for each valid block is normalized
to 1 BTC instead of the current 12.5 BTC. Moreover, we
calculate the reward as a probabilistic expectation for each
round. 

\item 
We do not consider unintentional forks. 
This assumption is reasonable because 
 the fork rates are negligible (the recent stale block rate is about 0.41\%~
\cite{gervais2016security}).
Because of this assumption, the reward for a miner is equal to the probability of 
finding a block by the miner for one round.
A period of finding a block by a miner has an exponential distribution with mean
inversely proportional to his computational power. 
Therefore, the probability of finding a block from a miner 
for one round is the same as his relative computational power.

\item  When a miner in a pool generates an FPoW, the manager
propagates a block corresponding to the FPoW and earns the reward. Then, the manager distributes the
reward to each miner in his pool in proportion to the miners'
submission shares for each round.
\end{enumerate}

\section{Attack Overview}
\label{Overview}

We describe a novel attack, called an FAW attack, combining 
selfish mining and a BWH attack. 
The core idea is that an attacker can split his computing power between innocent mining and infiltration mining, aiming at a target pool (as with a BWH attack).
However, when the attacker finds an FPoW as an infiltration miner, 
she deviates from the pattern of a BWH attack. In a BWH attack, the attacker drops 
the FPoW; in an FAW attack, she does not immediately propagate it to the pool manager, waiting instead for an external honest miner to publish theirs, 
at which point she propagates the FPoW to the manager hoping to cause a fork (similar to selfish mining). 
We present not only the FAW attack against one target pool but also 
a generalized FAW attack against multiple pools simultaneously. 
Finally, we present an \textit{FAW attack game} in which 
two pools attack each other via infiltration. 
The following are detailed descriptions of these FAW attack scenarios.

\vspace*{-1mm}

\subsection{One Target Pool}
\label{sec:one}
Considering an attacker who targets one open pool, the FAW attack
proceeds as follows. First, an attacker conducts both innocent
and infiltration mining by distributing her computational power to
join the target pool. If the attacker finds an FPoW through
innocent mining, she propagates it and earns a legitimate profit.
However, if the attacker finds an FPoW in the target
pool, she does not submit it immediately. After this, there are three
possible paths the attacker can take. 1) When she notices that other
miners, not participating in the target pool, propagate a valid block,
she immediately submits her FPoW to the manager of the
target pool, who propagates her FPoW to other Bitcoin nodes,
\textit{generating a fork in the Bitcoin network}. 2)
When an honest miner in the target pool finds an FPoW,
the attacker discards her FPoW. 3) When she finds another FPoW through
\textit{innocent mining}, she discards the FPoW generated by infiltration mining.
In summary,
the FAW attack generates intentional forks propagated by the
target pool, while the BWH attack \textit{never} does so. 
This detailed algorithm is Algorithm~\ref{FAWone} in 
Appendix~\ref{sec:algorithm}.

Based on this simple description, it is easy to see that the FAW attack is at least as profitable as the BWH attack. Note that the profit from the FAW attack is equal to that for the BWH attack in cases 2) and 3). In other words, additional profit comes from case 1). Suppose the attacker submits multiple FPoWs in case 1) over multiple rounds. If none of the FPoWs are chosen as the main chain, the profit from the FAW attack is equal to that from the BWH attack. If any of the attacker's FPoWs are chosen, the target pool receives a reward, which is distributed among miners including the infiltration miner. This gives additional profit to the attacker. 

Moreover, a manager's behavior can vary. If a manager
notices a valid block from outside the pool before the infiltration miner
submits her FPoW, an honest manager would discard the
FPoW generated by the infiltration miner.
However, if accepting the infiltration miner's FPoW is
more profitable (or would cause a smaller loss for the manager), a rational manager
may discard the FPoW from the outside instead. 
Otherwise, if an attacker propagates the withheld FPoW to the manager 
before the manager notices an external block propagation, 
the manager always selects the FPoW from the attacker regardless of his rational 
consideration.
We discuss
this rational behavior in more detail in Section~\ref{Discussion}.

\vspace*{-1mm}

\subsection{Multiple Target Pools}

An attacker can target multiple pools to generate a higher reward.
For simplicity, we first consider an FAW
attack executed against two pools (Pool$_{1}$ and Pool$_{2}$).
After the attacker joins the two target pools, she distributes her computational power for innocent mining and
infiltration mining between these pools. As in the single-pool case,
when the attacker finds an FPoW in Pool$_{1}$ or
Pool$_{2}$, she withholds it to generate a fork. However,
in this case, she may find two different
FPoWs, one for each pool, within a single round and withhold both. If
another honest miner propagates an FPoW, the attacker submits both FPoWs to both managers simultaneously. 
This behavior raises the winning probability of the infiltration miners' blocks 
in the fork by reducing propagation delay.
Therefore,
the attacker can make a fork that has two branches generated by herself 
and another found by an external honest miner, by letting two
target pools release two different valid blocks to the Bitcoin network
at the same time. 
When the attacker targets $n$ pools, she can execute the FAW attack as
above to generate a fork with $n+1$ branches.
The detailed algorithm is Algorithm~\ref{FAWmultiple} in Appendix
~\ref{sec:algorithm}.

\vspace*{-1mm}

\subsection{Pool vs Pool}

The activities of mining pools can be interpreted as a game in the Bitcoin system, with each pool choosing its strategy.
We consider the FAW attack as a strategy that pools can 
choose to earn higher rewards, 
meaning that an \textit{FAW attack game} can occur similarly to a BWH attack case~\cite{eyal2015miner}.
For simplicity, we assume that two pools, Pool$_1$ and Pool$_2$, play the game 
and all other miners are solo miners.

Pool$_1$ and Pool$_2$
first divide their own computational power into two parts for innocent and infiltration mining, and each pool infiltrates the other using its infiltration mining power.
While both conduct innocent and infiltration mining, if Pool$_1$ 
finds an FPoW in Pool$_2$ 
by infiltration mining, it withholds it.
After that, if Pool$_1$
generates an FPoW using innocent mining, it throws away its withheld FPoW
generated by infiltration mining, and the Pool$_1$ manager propagates the FPoW from its innocent mining.
The same action can be expected from Pool$_2$ with regard to Pool$_1$.
Otherwise, if someone from outside both pools broadcasts a valid block, 
the pools
generate a fork using their withheld FPoWs.
Therefore, a fork created under these conditions can include two or three branches (three branches might occur
if both Pool$_1$ and
Pool$_2$ have withheld FPoWs obtained from infiltration mining).
If both competing pools generate FPoWs through infiltration mining,
they select the FPoW generated from the opponent's infiltration mining 
for the main chain.
For example, the manager of Pool$_1$ selects the FPoW generated 
by infiltration mining of Pool$_2$ in Pool$_1$.

\section{FAW Attacks Against One Pool}
\label{Onepool}

In this section, we analyze the optimal behavior and maximum reward 
for an attacker theoretically and quantitatively when she targets one 
pool.
Our results show that the extra reward for an FAW attack is always equal
to or greater than that for a BWH attack.

\vspace*{-1mm}

\subsection{Theoretical Analysis}

We mathematically analyze our attack against one pool
and derive the optimal behavior of an attacker.
The relevant parameters are as follows:
\begin{itemize}
\item[$\alpha$:] Computational power of the attacker
\item[$\beta$:] Computational power of the victim pool
\item[$\tau$:] Attacker's Infiltration mining power as a proportion of $\alpha$
\item[$c$:] Probability that an attacker's FPoW through infiltration mining will be selected as the main chain
\end{itemize}
\noindent{}
The attacker uses computational power $(1-\tau)\alpha$
for innocent mining and $\tau\alpha$ for infiltration mining. 
Note that $\beta$ does not include the attacker's infiltration mining power in the victim pool. 
The parameter $c$ is a coefficient closely related to the topology of the 
Bitcoin network~\cite{miller2015discovering} and the attacker's network capability.
~\footnote{Network capability has been used in previous work~\cite{eyal2014majority,gervais2016security}, but $\gamma$ in those works is slightly different from $c$.}
We describe the parameter $c$ in detail in Section~\ref{Practicality}.

\begin{figure}[t]
\centering
\includegraphics[width=0.8\columnwidth]{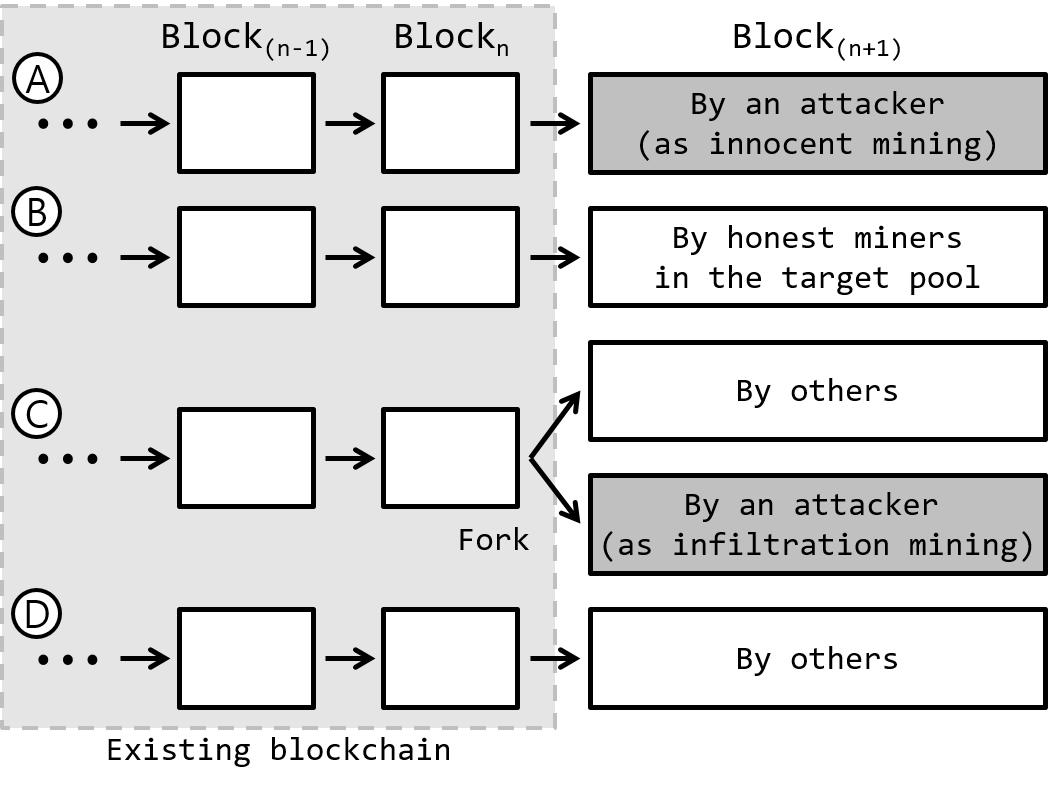}
\caption{Four cases of FAW attack results against one pool.
\circled{A} The attacker finds an FPoW through innocent mining, \circled{B} 
a miner other than the attacker in the target pool finds an FPoW,
\circled{C} the attacker finds an FPoW in the target pool and generates a fork, and \circled{D} someone else finds an FPoW, but she does not.
Blocks found by an attacker are displayed in dark gray. 
The attacker can earn rewards in cases \circled{A}, \circled{B}, and \circled{C}.
\vspace*{-3mm}}
\label{fig:newbwh}
\end{figure}

We can divide the attack results in each round into
four cases as shown in Fig.~\ref{fig:newbwh}. In the first case,
the attacker earns a reward through innocent mining.
Because she as an innocent miner should compete with others who have
total computational power $1-\tau\alpha$, the probability of the
first case is
$\frac{(1-\tau)\alpha}{1-\tau\alpha}$.
In the second case, the pool propagates an FPoW
found by an honest miner in the pool, with a probability of
$\frac{\beta}{1-\tau\alpha}$.
In the third case, when a valid block is found by an external honest miner 
(neither the attacker nor someone within the target pool), 
the attacker can generate a fork through the pool if she found and withheld an FPoW in advance. The probability is
$\tau\alpha\cdot\frac{1-\alpha-\beta}{1-\tau\alpha}$.
The final case occurs when a valid block is found by an external honest miner, 
but the attacker has \textit{not} found and withheld an FPoW. 
The probability of this case is $1-\alpha-\beta$.
As expected, the total probability of these four cases sums to 1.
Then, we can derive the FAW attacker's reward as follows.

\begin{theorem}
An FAW attacker can earn 
  \begin{equation}
\label{eq:one}
    R_a(\tau)=\frac{(1-\tau)\alpha}{1-\tau\alpha}
+\left(\frac{\beta}{1-\tau\alpha}
+c\tau\alpha\cdot\frac{1-\alpha-\beta}{1-\tau\alpha}\right)\cdot\frac{\tau\alpha}{\beta+\tau\alpha}.
  \end{equation}
The reward is maximized when the 
optimal $\tau$ value, $\overline{\tau}$, is

\begin{equation}
\label{eq:op_tau}
\resizebox{\hsize}{!}{%
  $   \frac{(1-\alpha)(1-c)\beta+\beta^2c-\beta\sqrt{(1-\alpha-\beta)^2c^2+((1-\alpha-\beta)(\alpha\beta+\alpha-2))c-\alpha(1+\beta)+1}}
  {\alpha(1-\alpha-\beta)(c(1-\beta)-1)} $
}
\end{equation}
\end{theorem}

\begin{figure*}[ht]
\small
\begin{equation}
\label{eq:gamma_der}
\frac{\partial R_a}{\partial \gamma} = \frac{\alpha^2\beta^2+(((2\alpha^2-2\alpha^3)\beta
-2\alpha^2\beta^2)c + (2\alpha^3-2\alpha^2)\beta)\gamma+
((\alpha^3-\alpha^4+(\alpha^4-2\alpha^3)\beta+\alpha^3\beta^2)c+\alpha^3\beta+\alpha^4
-\alpha^3)\gamma^2}
{\alpha^4 \gamma^4 +
(2\alpha^4\beta-2\alpha^3)\gamma^3+(\alpha^2-4\alpha^2\beta+\alpha^2\beta^2)\gamma^2+
(2\alpha\beta-2\alpha\beta^2)\gamma+\beta^2}=0
\end{equation}
\end{figure*}

\begin{myproof}[Proof Sketch]
Because an attacker works as both an innocent and infiltration miner, she is rewarded in both roles.
Her reward from innocent mining is 
\begin{equation}
\frac{(1-\tau)\alpha}{1-\tau\alpha}\notag
\end{equation}
(case \circled{A} in Fig.~\ref{fig:newbwh}).
To derive her reward from infiltration mining, we first describe the reward for the pool to which the infiltration miner belongs.
The pool can earn a profit in two cases: when an honest miner in the pool generates an FPoW (case \circled{B}),
and when the attacker successfully generates a fork and her FPoW is selected as the main chain (case \circled{C}).
In case \circled{B}, the pool earns the reward $\frac{\beta}{1-\tau\alpha}$. 
In case \circled{C}, the reward for the pool is
$c\tau\alpha\cdot\frac{1-\alpha-\beta}{1-\tau\alpha}$ 
through the fork generated by the attacker. 
Therefore, the pool can earn the reward 
\begin{equation}
\frac{\beta}{1-\tau\alpha}
+c\tau\alpha\cdot\frac{1-\alpha-\beta}{1-\tau\alpha}.\notag
\end{equation}

Then the pool manager pays a reward proportional to the attacker's submitted
(both full and partial) PoWs, and the attacker's estimated contribution from 
the pool manager is $\frac{\tau\alpha}{\beta+\tau\alpha}$.
As a result, the attacker's reward $R_a$ can be expressed with Eq.~\eqref{eq:one}.
The attacker's reward $R_a$ is a function of $\tau$, and we 
can find the value of $\tau$ that maximizes $R_a$ by solving 
Eq.~\eqref{eq:gamma_der}. 
We call this value of $\tau$ as $\overline{\tau}$.
Finally, $\overline{\tau}$ is expressed in Eq.~\eqref{eq:op_tau}.
\end{myproof}

\vspace*{-1mm}

According to the theorem above, an attacker should distribute her infiltration 
mining power as an optimal portion $\overline{\tau}$ of her total power 
to earn the maximum reward.
Additionally, an FAW attack with optimal $\overline{\tau}$ satisfies the following theorem.

\begin{theorem}
\label{th:main}
An FAW attack is always more profitable than honest mining, 
and the reward from an FAW attack has a lower bound defined by the reward from a BWH attack.  
\end{theorem}

\begin{myproof}[Proof Sketch]
We show that the attacker's reward, 
$R_a(\overline{\tau})$, is always greater than the
honest miner's reward $\alpha$.
First, the reward $R_a$ when $c=0$ is equal to the reward from the BWH attack
since a case where the FAW attacker receives zero reward due to a
fork is equivalent to the BWH attack. Luu et al.~\cite{luu2015power} proved 
that the BWH attacker's reward can always be larger than $\alpha$ when a proper 
value of $\tau$ is chosen. 
Furthermore, $R_a$ is an increasing function of $c$.
As a result, an FAW attack produces an extra reward
regardless of the attacker's computational power, as in the BWH attack.
\end{myproof}

\vspace*{-1mm}

\vspace*{-1mm}

Theorem~\ref{th:main} states as mentioned intuitively in Section~\ref{sec:one} that the FAW attack is at
least as profitable as the BWH attack.
Note that $\overline{\tau}$ depends on a
constant $c$, related to network topology~\cite{miller2015discovering, bitnodes}.
To maximize reward,
an attacker must know the value of $c$. For now, we assume that $c$ is given to the
attacker, but learning $c$ is not easy in practice. Nevertheless,
we show in Section~\ref{Practicality}
that the FAW attack still improves upon the BWH attack even when $c$ is unknown.

Next, our focus moves to the target pool's loss. Through the following theorem, 
it is shown that the target pool's reward after the FAW attack is always
smaller than that it would be without, though incentives do exist for the target pool manager
to propagate the FPoW found by the attacker's infiltration mining even if 
he notices the FPoW is stale.

\begin{theorem}
\label{thm_victim}
   The reward for the target pool is 
    $R_{p}=\frac{\beta}{1-\tau\alpha}+c\tau\alpha\frac{1-\alpha-\beta}{1-\tau\alpha}$, 
and this is always less than $\beta+\tau\alpha$, which is 
the target pool's reward without the FAW attack.
Additionally, reward $R_{p}$ is an increasing function of $c$.
\end{theorem}

\begin{myproof}[Proof Sketch]
The target pool earns the reward $\frac{\beta}{1-\tau\alpha}$ in 
case~\circled{B} and $c\tau\alpha\frac{1-\alpha-\beta}{1-\tau\alpha}$ in case~\circled{C}.
Therefore, $R_p$ can be expressed as 
\begin{equation}
\frac{\beta}{1-\tau\alpha}+c\tau\alpha\frac{1-\alpha-\beta}{1-\tau\alpha},\notag
\end{equation}
and $R_p$ is a linear function of $c$ with positive coefficient. 
This means that $R_p$ is an increasing function of $c$. 
Finally, we show that $R_p$ is always less than $\beta+\tau\alpha$. 
\begin{align}
R_{p}&=\frac{\beta}{1-\tau\alpha}+c\tau\alpha\frac{1-\alpha-\beta}{1-\tau\alpha}\notag
\\
&\leq
\frac{\beta}{1-\tau\alpha}+\tau\alpha\frac{1-\alpha-\beta}{1-\tau\alpha}
~~~\text{when }c=1 \notag\\
&< \beta +\tau\alpha~~~~~~~~~~~~~~~~~\text{when } 0 \leq
\tau < 1 \notag\\
\Longleftrightarrow &\,\,\,\beta+\tau\alpha (1-\alpha-\beta) < (\beta+\tau\alpha)
(1-\tau\alpha)\notag\\
\Longleftrightarrow &\,\,\,\beta+\tau\alpha-\tau\alpha^2-\tau\alpha\beta
<\beta-\beta\tau\alpha+\tau\alpha-\tau^2\alpha^2\notag\\
\Longleftrightarrow &\,\,\,\tau^2\alpha^2  < \tau\alpha^2\notag \\
\Longleftrightarrow &\,\,\,\tau  < 1\notag 
\label{eq:pool_r}
\end{align}
Because $\tau$ is less than 1, $R_p$ is always less than $\beta+\tau\alpha$.
\end{myproof}
Note that the target pool's loss decreases as $c$ increases.
Therefore, the pool manager should try to increase $c$ to reduce loss.
Thus, he should propagate his FPoWs as fast as
he can, which incidentally also increases the attacker's extra
reward ($R_a$ in Eq.~\eqref{eq:one}).

\subsection{Quantitative Analysis}
\label{subsec:one_quant}
\begin{figure}[ht]
\centering{
\subfloat[The RER (\%) of an attacker, $R^{'}_{a}$, according to target pool size $\beta$ and network capability $c$
when the attacker's computational power $\alpha$ is 0.2. \vspace*{-2mm} ]{
\includegraphics[width=0.5\columnwidth]{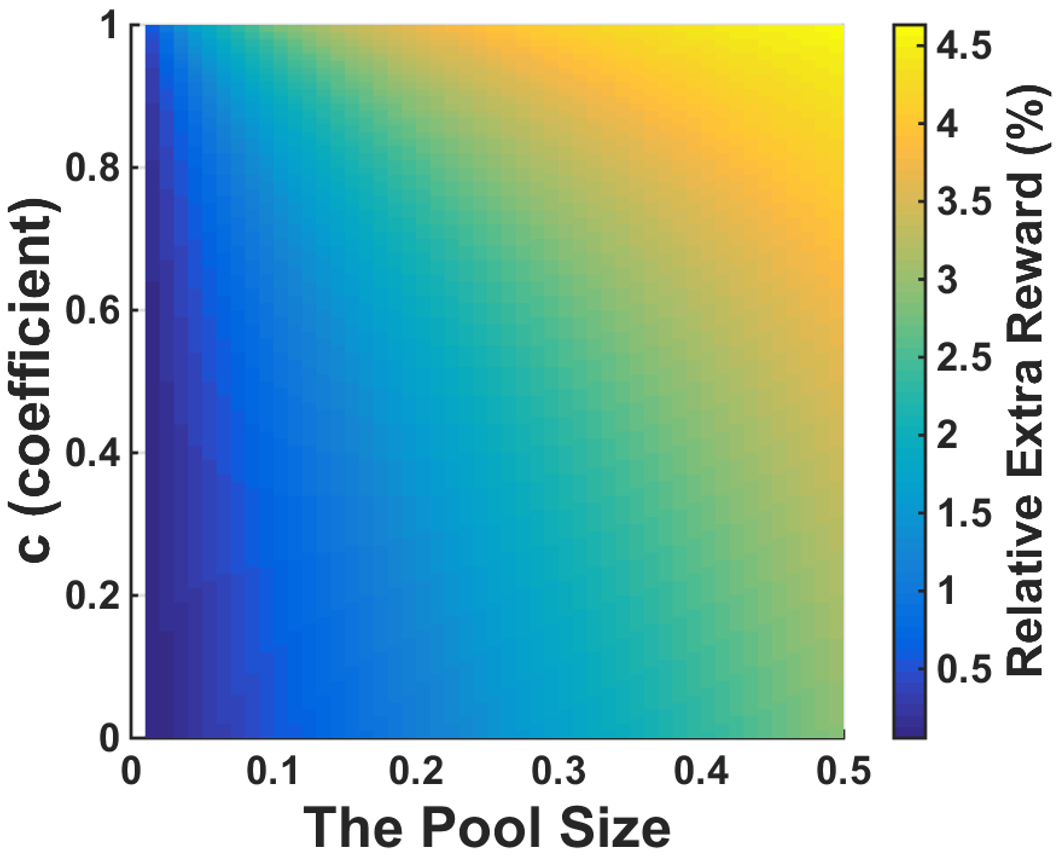}
\label{fig:one2}
}
\subfloat[The RER (\%) of a target pool, $R^{'}_{p}$, according to $\beta$ and $c$
when the attacker's computational power $\alpha$ is 0.2. Negative RER
means loss.\vspace*{-2mm} ]{
\includegraphics[width=0.5\columnwidth]{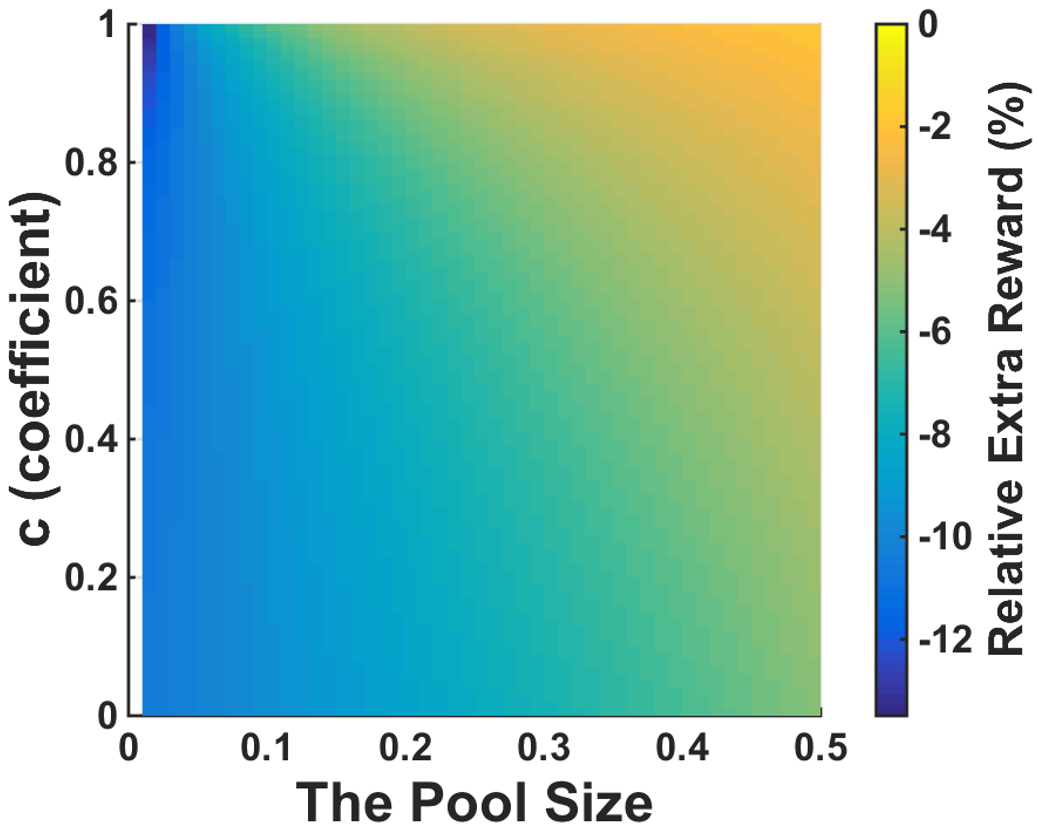}
\label{fig:one2_p}
}}
\caption{Quantitative analysis results for the FAW attack against one
pool. When $c$ increases, attacker's reward increases and the target
pool's loss decreases.\vspace*{-3mm}}
\label{fig:one}
\end{figure}

In this section we consider a specific case: an attacker
with computational power 0.2, who executes an FAW attack
against one pool. 
We define the \textit{relative extra reward} (RER) gained with respect to the
reward $R_h$ of an honest miner, which is equivalent to his computational
power. The RER $R^{'}_{a}$ of an attacker can be expressed as
\begin{equation}
R^{'}_{a}=\frac{R_{a}-R_{h}}{R_{h}}.\notag
\end{equation}
In the same manner, the RER of the target pool is 
\begin{equation}
R^{'}_{p}=\frac{R_{p}-R_{h}}{R_{h}}. \notag
\end{equation}
(A negative value indicates a loss.)
Figs.~\ref{fig:one2} and~\ref{fig:one2_p} show the RER of the
attacker and a victim pool, respectively, given terms $c$ and $\beta$
when the attacker's computational power $\alpha$ is 0.2.

Fig.~\ref{fig:one2} demonstrates that
an attacker can earn an extra reward regardless of $c$ or the target pool size $\beta$.
Therefore, an attacker should always run the FAW attack to increase her own reward.
Moreover, increasing $c$ provides an even greater extra reward.
As noted previously, when $c$ is zero, the RERs for BWH and FAW attacks are the same.
Therefore, the extra reward for the FAW attacker is always lower bounded 
by that for the BWH attacker.
Thus, the FAW attack improves on the BWH attack in all cases.

Conversely, Fig.~\ref{fig:one2_p} confirms
that a target pool always suffers a loss in the presence of an
attacker. (A negative extra reward indicates a loss.)
However, the loss of the target pool decreases as the value of $c$ increases 
because when the FPoW generated by an attacker in the target pool is selected
as the main chain, the target pool also earns a reward for the block.

\subsection{Simulation Results}

To verify the theoretical analysis developed, we simulated an FAW attack against one pool with a computational power of 0.2, using a Monte Carlo method over $10^9$ rounds, with an upper bound of
$10^{-4}$ for error. Table~\ref{tab:com}
shows the attacker's RER (\%) according to her computational power $\alpha$ and $c$ when $\beta$ is 0.2.
She can always earn the extra reward by executing the FAW attack, and 
her extra reward is equal to or greater than that for the BWH attacker.

\begin{table}[ht]
\caption{The RER (\%) of an attacker when target pool size $\beta$ is 0.2.
The value $a\,(b)$ gives RERs based on theoretical analysis and simulation, respectively. \vspace*{-1mm}}
\begin{center}
\begin{tabular}{|c||c|c|c|c|c|c|c|c|}
\hline
\begin{tabular}[c]{@{}c@{}}\diagbox{\textbf{c}}{$\alpha$}\end{tabular} & \textbf{0.1} & \textbf{0.2} & \textbf{0.3} & \textbf{0.4} \\ \hline\hline
\textbf{0} & \begin{tabular}[c]{@{}c@{}}0.53 (0.53)\end{tabular} & \begin{tabular}[c]{@{}c@{}}1.14 (1.14)\end{tabular} & \begin{tabular}[c]{@{}c@{}}1.85 (1.85)\end{tabular} & \begin{tabular}[c]{@{}c@{}}2.70 (2.70)\end{tabular} \\ \hline
\textbf{0.25} & \begin{tabular}[c]{@{}c@{}}0.65 (0.67)\end{tabular} & \begin{tabular}[c]{@{}c@{}}1.38 (1.38)\end{tabular} & \begin{tabular}[c]{@{}c@{}}2.20 (2.20)\end{tabular} & \begin{tabular}[c]{@{}c@{}}3.1 (3.13)\end{tabular} \\ \hline
\textbf{0.5} & \begin{tabular}[c]{@{}c@{}}0.85 (0.85)\end{tabular} & \begin{tabular}[c]{@{}c@{}}1.74 (1.74)\end{tabular} & \begin{tabular}[c]{@{}c@{}}2.70 (2.70)\end{tabular} & \begin{tabular}[c]{@{}c@{}}3.75 (3.75)\end{tabular} \\ \hline
\textbf{0.75} & \begin{tabular}[c]{@{}c@{}}1.21 (1.22)\end{tabular} & \begin{tabular}[c]{@{}c@{}}2.37 (2.37)\end{tabular} & \begin{tabular}[c]{@{}c@{}}3.52 (3.52)\end{tabular} & \begin{tabular}[c]{@{}c@{}}4.69 (4.70)\end{tabular} \\ \hline
\textbf{1} & \begin{tabular}[c]{@{}c@{}}2.12 (2.12)\end{tabular} & \begin{tabular}[c]{@{}c@{}}3.75 (3.75)\end{tabular} & \begin{tabular}[c]{@{}c@{}}5.13 (5.13)\end{tabular} & \begin{tabular}[c]{@{}c@{}}6.37 (6.36)\end{tabular} \\ \hline
\end{tabular}

\end{center}
\label{tab:com}
\end{table}

\section{FAW Attack Against Multiple Pools}
\label{multipool}

An attacker should maximize her reward by
targeting $n$ pools simultaneously. She 
distributes her infiltration power among $n$ pools and can 
find at most $n$ FPoWs, one for each of $n$ different pools 
within a given round, so she can generate
a fork that has a maximum of $n+1$ branches. 
In this section, we analyze
this scenario theoretically and quantitatively. 
Unless otherwise stated, we describe the $n$-pool
attack using an example where $n=2$ for ease of exposition.

\begin{figure}[t]
\centering
\includegraphics[width=0.7\columnwidth]{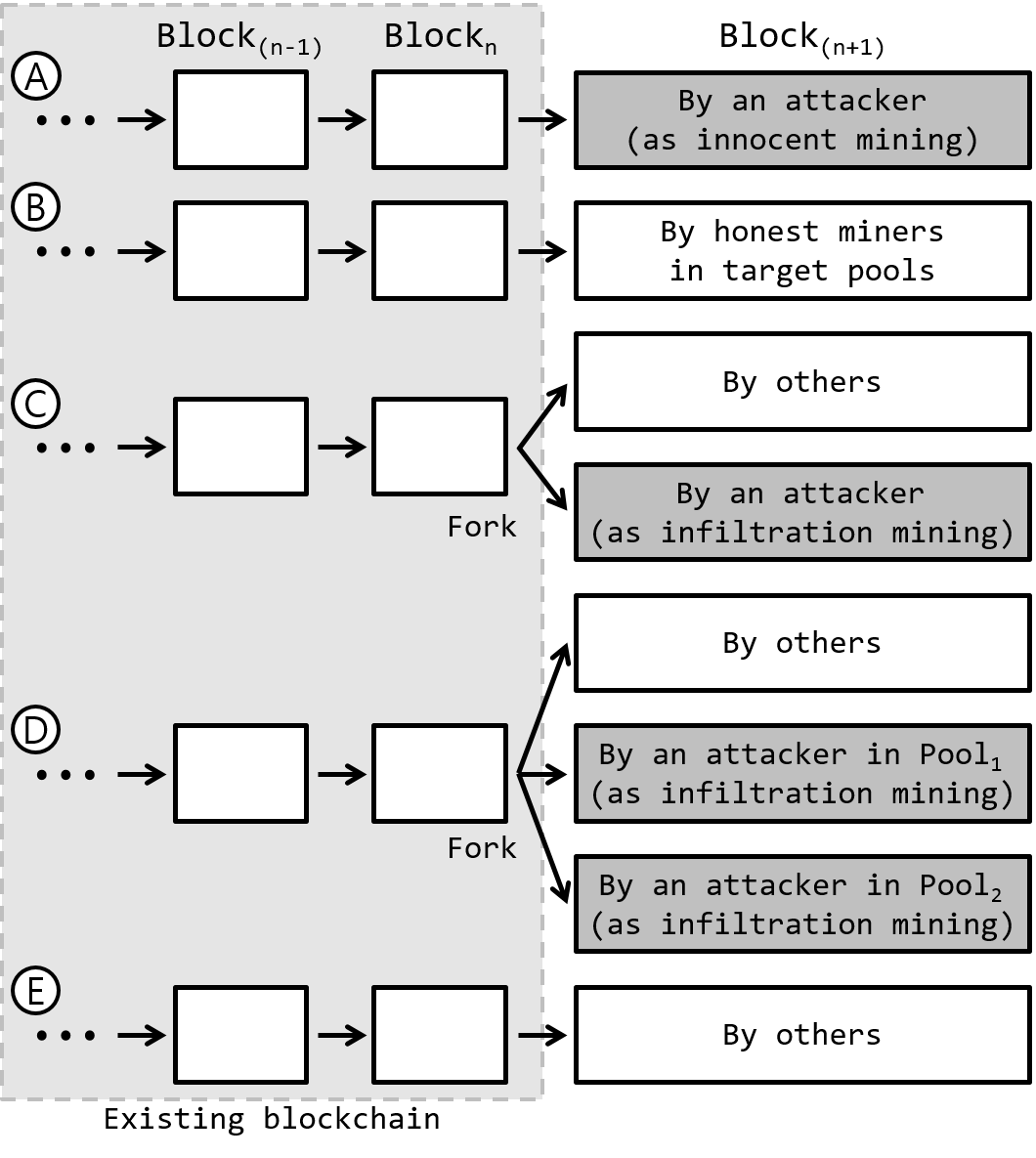}
\caption{Five cases of FAW attack results against multiple pools.
\circled{A} An attacker finds an FPoW through innocent mining, 
\circled{B} another miner in the target pool finds an FPoW,
\circled{C} the attacker finds an FPoW in one target pool and generates a fork, \circled{D} the attacker finds an FPoW in multiple target pools
and generates a fork, and \circled{E} someone else finds an FPoW. 
The attacker can earn rewards
in cases \circled{A}, \circled{B}, \circled{C}, and \circled{D}.}
\label{fig:multi newbwh}
\end{figure}

\subsection{Theoretical Analysis}

Let the computational power of an attacker be $\alpha$ 
and the power of Pool$_{1}$ and Pool$_{2}$ be $\beta_1$ and $\beta_2$, respectively.
The attacker distributes her computational power into $\tau_1$ and $\tau_2$ fractions for infiltration mining in Pool$_{1}$ and Pool$_{2}$, respectively.
When an attacker withholds an FPoW in Pool$_i$ only, and 
an external honest miner releases a valid block (Case \circled{C} in Fig.~\ref{fig:multi newbwh}), 
the variable $c_i^{(1)}$ represents the probability 
that the FPoW of the infiltration miner in Pool$_i$ will be 
selected as the main chain.
Variable $c_i^{(2)}$ is the probability that the FPoW found by
her infiltration mining in Pool$_i$ will be selected as the main chain 
among three branches if she withholds FPoWs from both pools when 
an external honest miner propagates a valid block (Case \circled{D} in Fig.~\ref{fig:multi newbwh}).
Therefore, the sum of $c_1^{(2)}$
and $c_2^{(2)}$ must be less than or equal to 1.
Then we can derive her reward $R_a$ as follows.

\begin{theorem}
\label{twopools}
When the FAW attacker executes the FAW attack against Pool$_1$ and
Pool$_2$, she can earn reward $R_a$ as
\begin{equation}
\label{eq:two}
  \footnotesize
  \begin{split}
        & \frac{(1-\tau_1-\tau_2)\alpha}{1-(\tau_1+\tau_2)\alpha} +
        \sum_{i=1,2} \Bigg\{\frac{\tau_i\alpha}{\beta_i+\tau_i\alpha} \Bigg(\frac{\beta_i}{1-(\tau_1+\tau_2)\alpha} \\
        &\hspace{5mm}+ c_i^{(1)}\tau_i\alpha\frac{1-\alpha-\beta_1-\beta_2}{1-\tau_i\alpha} 
        + c_i^{(2)}\sum_{j}\{\tau_j\alpha\frac{\tau_{\neg j}\alpha}
        {1-\tau_i\alpha}\}\frac{1-\alpha-\beta_1-\beta_2}{1-(\tau_1+\tau_2)\alpha}
        \Bigg)
        \Bigg\}
  \end{split}
\end{equation}
\end{theorem}

\begin{myproof}[Proof Sketch]
The total reward for the attacker is composed of rewards from innocent mining 
and infiltration mining in Pool$_{1}$ and Pool$_{2}$.
The reward from innocent mining (case \circled{A} in Fig.~\ref{fig:multi newbwh}) is
\begin{equation}
\frac{(1-\tau_1-\tau_2)\alpha}{1-(\tau_1+\tau_2)\alpha}.\notag 
\end{equation}
Prior to deriving the infiltration mining part of the attacker's reward 
from Pool$_{1}$ and Pool$_{2}$, 
we derive the total reward for each target pool.
When an FPoW is found by an honest miner in the target pools, (case \circled{B} in Fig.~\ref{fig:multi newbwh}), target Pool$_i$ can earn
\begin{equation}
\frac{\beta_i}{1-(\tau_1+\tau_2)\alpha}.\notag
\end{equation}
Next, if the attacker generates an intentional fork with two branches (case \circled{C} in Fig.~\ref{fig:multi newbwh}), and the attacker's FPoW is selected as the main chain, 
Pool$_i$ can earn

\begin{equation}
\footnotesize
c_i^{(1)}\tau_i\alpha\frac{1-\alpha-\beta_1-\beta_2}{1-\tau_i\alpha}.\notag
\end{equation}
Finally, we consider case \circled{D} when the attacker generates an intentional fork with three branches and the attacker's FPoW is selected as the main chain.
This case means that she finds two FPoWs in both Pool$_1$ and Pool$_2$ within 
one round, respectively.
If the attacker first finds an FPoW in Pool$_1$ and the FPoW is selected as the 
main chain, Pool$_1$ can earn the reward 
\begin{equation}
c_1^{(2)}\tau_1\alpha\frac{\tau_2\alpha}{1-\tau_1\alpha}\frac{1-\alpha-\beta_1-\beta_2}{1-(\tau_1+\tau_2)\alpha}.\notag
\end{equation}
Otherwise if the attacker finds another FPoW in Pool$_1$ after she finds an
FPoW in Pool$_2$ and the attacker's FPoW in Pool$_1$ is selected as the main chain,
then Pool$_1$ can earn the reward
\begin{equation}
c_1^{(2)}\tau_2\alpha\frac{\tau_1\alpha}{1-\tau_2\alpha}\frac{1-\alpha-\beta_1-\beta_2}{1-(\tau_1+\tau_2)\alpha}\notag
\end{equation}
As a result, Pool$_i$ can earn 
\begin{equation}
c_i^{(2)}\sum_{j=1,2}\tau_j\alpha\frac{\tau_{\neg j}\alpha}{1-\tau_j\alpha}\frac{1-\alpha-\beta_1-\beta_2}{1-(\tau_1+\tau_2)\alpha}\notag
\end{equation}
through case \circled{D}, and the total reward of Pool$_i$ is 
\begin{equation}
\resizebox{\hsize}{!}{%
$\frac{\beta_i}{1-(\tau_1+\tau_2)\alpha} + c_i^{(1)}\tau_i\alpha\frac{1-\alpha-\beta_1-\beta_2}{1-\tau_i\alpha} + c_i^{(2)}\sum_{j}\{\tau_j\alpha\frac{\tau_{\neg j}\alpha}{1-\tau_i\alpha}\}\frac{1-\alpha-\beta_1-\beta_2}{1-(\tau_1+\tau_2)\alpha}$.\notag
}
\end{equation}
Then the reward for the attacker from Pool$_i$ is a fraction
$\frac{\tau_{i}\alpha}{\beta_{i}+\tau_i\alpha}$ of the total
reward for Pool$_i$. Therefore, considering all cases, the total reward for the attacker, $R_a$, can be derived by
Eq.~(\ref{eq:two}).
\end{myproof}

\vspace*{-1mm}

Below, we expand to the FAW attack targeting $n$ pools, computing the attacker's reward $R_a$. 
The theorem can be proven in a similar way as Theorem~\ref{twopools}.

\begin{theorem}
Generalization for $n$ pools,
where the computational power of target Pool$_i$ is $\beta_i$ and 
the fraction of the attacker's power devoted to the pool is $\tau_i$.
The total reward for the attacker, $R_a$, is 
\begin{equation}
  \label{eq:npools}
  \footnotesize
  \begin{split}
      R_a &= \frac{(1-\tau)\alpha}{1-\tau\alpha} + \sum_{i=1}^{n} \Bigg[ \frac{\tau_i\alpha}{\beta_i+\tau_i\alpha} \Bigg(\frac{\beta_i}{1-\tau\alpha} \\
         &+  \sum_{k=1}^{n}\Bigg\{(1-\alpha-\beta)
              \sum_{\mathcal{P}_{k,i}\in\mathcal{P}} c_{\mathrm{Im}(\mathcal{P}_{k,i})}(i)\prod_{t=1}
         ^{k}\frac{\tau_{\mathcal{P}_{k,i}(t)}\alpha}{1-\sum_{d=1}^{t}\tau_{\mathcal{P}_{k,i}(d)\alpha}}
         \Bigg\}
       \Bigg) 
     \Bigg],
  \end{split}
\end{equation}
when attacking $n$ pools with the following conditions hold:
$\tau = \sum_{i=1}^{n}\tau_i$, $\beta = \sum_{i=1}^{n}\beta_i$, 
$\mathcal{P}_{k,i}$ is a one-to-one function from $\{1,2,...,k\}$ to
$\{1,2, ...,n\}$, where an image of $\mathcal{P}_{k,i}$ (i.e.,
$\mathrm{Im}(\mathcal{P}_{k,i})$) must include $i$, and
$c_{\mathrm{Im}(\mathcal{P}_{k,i})}(i)$ is the probability
that the attacker's FPoW in Pool$_i$ will be selected as the main chain when she finds
one FPoW in each of $k$ pools.
\end{theorem}

\begin{myproof}[Proof Sketch]
First, the attacker can earn the reward $\frac{(1-\tau)\alpha}{1-\tau\alpha}$ from 
innocent mining. When an honest miner in Pool$_i$ finds an FPoW, the attacker 
can earn the reward 
\begin{equation}
\frac{\beta_i}{1-\tau\alpha}\cdot 
\frac{\tau_i \alpha}{\beta_i+\tau_i\alpha}.\notag
\end{equation}
Next, we consider the case when she generates forks with $k$ branches. 
If she finds and withholds an FPoW in each of $k$ pools including Pool$_i$, and 
the FPoW from Pool$_i$ is selected as the main chain,
Pool$_i$ as well as the attacker can earn rewards. 
From this case, Pool$_i$ can earn the reward 
\begin{small}
\begin{equation}
(1-\alpha-\beta)
              \sum_{\mathcal{P}_{k,i}\in\mathcal{P}} c_{\mathrm{Im}(\mathcal{P}_{k,i})}(i)\prod_{t=1}
         ^{k}\frac{\tau_{\mathcal{P}_{k,i}(t)}\alpha}{1-\sum_{d=1}^{t}\tau_{\mathcal{P}_{k,i}(d)\alpha}}.\notag
\end{equation}
\end{small} 
Then the attacker earns a
$\frac{\tau_i \alpha}{\beta_i+\tau_i\alpha}$ portion of the above reward.
Finally, when considering all values of $k$ and $i$, the total reward for the attacker is Eq.~\eqref{eq:npools}.
\end{myproof}

Eq.~\eqref{eq:npools} is a function of $\tau_i$
($i=1, \ldots , n$); therefore, an attacker can maximize her RER $R_a^{'}$ depending on the value of $\tau_i$ ($i=1, \ldots , n$).
Moreover, the total reward for each target Pool$_i$ increases as $c_{\mathrm{Im}(\mathcal{P}_{k,i})}(i)$ increases.
Therefore, to reduce loss, target pool managers should try to increase
$c_{\mathrm{Im}(\mathcal{P}_{k,i})}(i)$, which in turn increases the attacker's extra reward.

\subsection{Quantitative Analysis}

\begin{figure}[ht]
\centering
\includegraphics[width=\columnwidth]{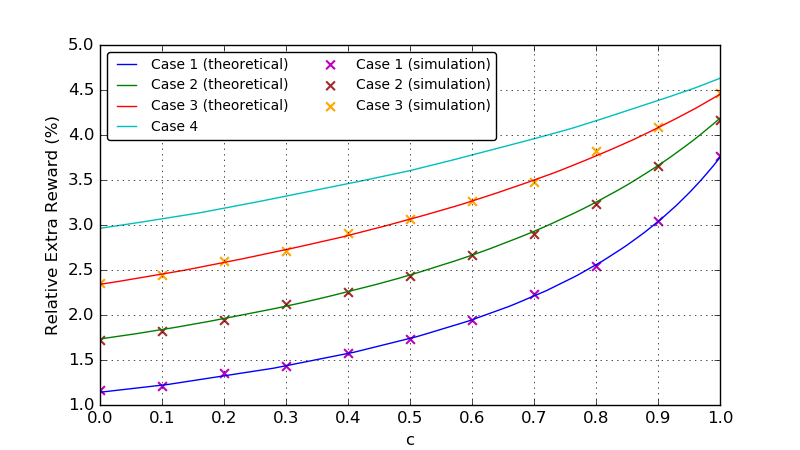}
\caption{Rewards for an FAW attacker against
two pools when her computational power is $\alpha =0.2$. Cases 1, 2, and 3 represent two
target pools with computational power $(\beta_1, \beta_2)$ equal to (0.1,
0.1), (0.2, 0.1), and (0.3, 0.1), respectively. Case 4 represents when F2Pool executes the FAW attack against all open pools
in Table~\ref{tab:pool}. Theoretical analysis result matches with simulation results approximately.}
\label{fig:multi reward}
\end{figure}

Seven parameters are used to represent a two-pool attack, 
which
determine the attacker's RER:
$\alpha, \beta_i, c^{(j)}_i$ for $i=1, 2$ and $j=1, 2$.
For simplicity, 
we make the following assumptions:
first, the attacker's computation power, $\alpha$, is assumed to be 0.2.
Three cases for the two pools' power:
cases 1, 2, and 3 have ($\beta_1$, $\beta_2$) equal to (0.1, 0.1), (0.2, 0.1), and (0.3, 0.1), respectively.
We also assume $c_{\mathrm{Im}(\mathcal{P}_{k,i})}(i)=\frac{c}{k}$ 
where $c$ ranges from 0 to 1.
Fig.~\ref{fig:multi reward} shows the attacker's RERs (\%) for various values of $c$. As expected, as $c$
increases, RER also increases. Furthermore, when
the total computational power of the two target pools increases, 
RER also increases.

As an additional case (case 4), we also analyzed the FAW attacker's RER, taking an approximate computational power distribution from
the current Bitcoin network as shown in Table~\ref{tab:pool}, obtained from~\cite{hashdist}.
Assume that F2Pool executes the FAW attack against four other open pools.
In this case, AntPool, BTCC Pool, BW.com, and BitFury correspond to Pool$_1$, Pool$_2$, Pool$_3$, and Pool$_4$, respectively.
Because of the symmetry between three pools, optimal
values for infiltration mining power as a portion of the
attacker's computational power for each target pool (i.e., $\tau_2$,
$\tau_3$, and $\tau_4$) are the same.


The RER for an attacker in case 4 is also shown in Fig.~\ref{fig:multi reward}.
Considering the current pool distribution shown in Table~\ref{tab:pool}, the BWH attack gives the attacker an RER
of $2.96\%$, but she can earn a maximum RER of $4.63\%$ with the FAW attack.
Therefore, the FAW attack gives her an extra reward of $56.24\%$ more than that the BWH attack.

\begin{table}[ht]
\vspace*{-2mm}
\centering
\caption{Approximate Bitcoin power distribution~\cite{hashdist},
including closed pools and solo miners marked as Unknown. }

\resizebox{1\columnwidth}{!}{%
\begin{tabular}{|c|c||c|c|}
\hline
Owner & Computational Power & Owner & Computational Power \\
  \hline\hline
Unknown & 30\% & BTCC Pool & 10\% \\ \hline
F2Pool & 20\% & BW.com & 10\% \\ \hline
AntPool & 20\% & BitFury & 10\% \\ \hline
\end{tabular}
} 
\label{tab:pool}
\end{table}

\subsection{Simulation Results}

To verify the accuracy of this analysis, we implemented a Monte Carlo simulator in
Python to simulate an FAW attack 
against the two pools in cases 1, 2, and 3 in Fig.~\ref{fig:multi reward}.
The $\times$-marks show simulation results for $10^8$ rounds, confirming the calculations.

 \section{Two-Pool FAW Attack Game}
\label{Game}

As described in Section~\ref{Overview}, pools can execute FAW attacks 
against each other as well. 
We model a simultaneous game between two players, Pool$_1$ and Pool$_2$.
We know that compliance with Bitcoin protocol by both players is not a
Nash equilibrium, because the FAW attacker can earn extra rewards as
discussed in Sections~\ref{Onepool} and~\ref{multipool}. 
In this section, we prove and derive the following result in the Nash equilibrium.
In the case of an FAW attack, 1) the miner's dilemma no longer applies, and 
2) the game outcome is based on pool size, where
the larger pool wins the game.
Note that while the game is generalizable to $n$ pools,
we leave an exact analysis for future work.
Before analyzing the two-pool FAW attack game,
we define \textit{the winning condition}
as earning an extra reward.
By this definition, the game outcome indicates either a single winner, or
no winner (as in the miner's dilemma).

\begin{figure}[t]
\centering
\includegraphics[width=0.8\columnwidth]{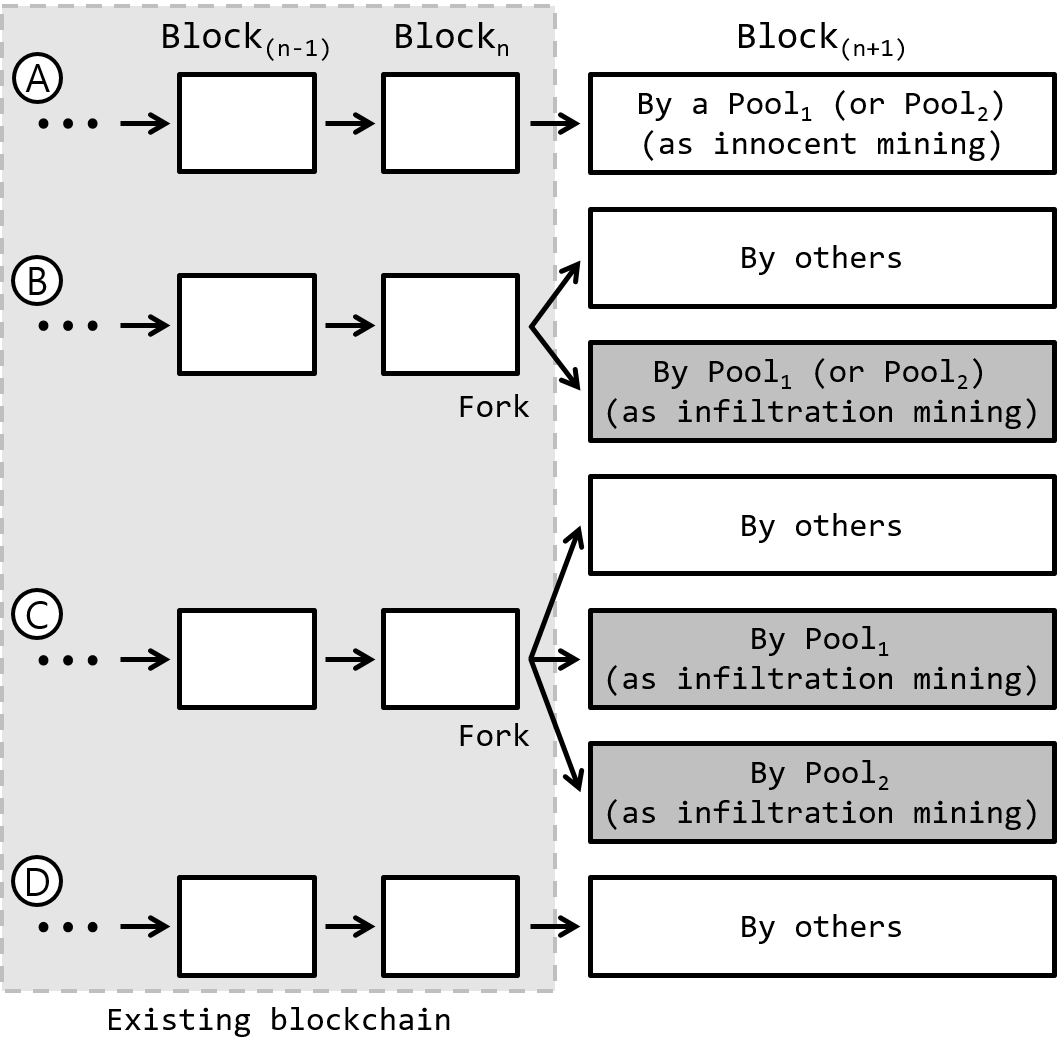}
\caption{Four cases of the two-pool FAW attack game.
\circled{A} Pool$_1$ (or Pool$_2$) finds an FPoW by innocent mining, 
\circled{B} Pool$_1$ (or Pool$_2$) finds an FPoW using infiltration mining and generates a fork,
\circled{C} Pool$_1$ and Pool$_2$ both find an FPoW in the opponent pool through infiltration mining and generate a fork, and 
\circled{D} someone else finds an FPoW. 
Each pool can earn a reward in cases \circled{A}, \circled{B}, and \circled{C}. \vspace*{-3mm}
}
\label{fig:game newbwh}
\end{figure}

\subsection{Theoretical Analysis of the Game}
\label{subsec:game-theory}
Parameters for the analysis of the FAW attack game are defined as
below for $i=1,2$.

\begin{itemize}
  \item[$\alpha_i$:] Computational power of Pool$_i$
  \item[$f_i$:] Infiltration mining power of Pool$_i$, i.e., $f_i = \tau_i\alpha_i$
\end{itemize}

When both rational players choose the FAW attack as a strategy, the
players' rewards are as follows.

\begin{theorem}
In the FAW attack game between two pools, the rewards $R_1$ of Pool$_1$ and $R_2$ of Pool$_2$ are:

\begin{equation}
\label{eq:game_pool1}
\resizebox{0.91\hsize}{!}{%
  $ R_1 =\frac{\alpha_1-f_1}{1-f_1-f_2}
  + c_2f_2\frac{1-\alpha_1-\alpha_2}{1-f_2} 
  +
  c_2^{'}f_1f_2(\frac{1}{1-f_1}+\frac{1}{1-f_2})\frac{1-\alpha_1-\alpha_2}{1-f_1-f_2}
  + R_2\frac{f_1}{\alpha_2+f_1}$
}
\end{equation}
\begin{equation}
\label{eq:game_pool2}
\resizebox{0.91\hsize}{!}{%
  $ R_2=\frac{\alpha_2-f_2}{1-f_1-f_2}
  + c_1f_1\frac{1-\alpha_1-\alpha_2}{1-f_1} 
  +c_1^{'}f_1f_2(\frac{1}{1-f_1}+\frac{1}{1-f_2})\frac{1-\alpha_1-\alpha_2}{1-f_1-f_2}
  + R_1\frac{f_2}{\alpha_1+f_2}
  $
}
\end{equation}
\end{theorem}

\begin{myproof}[Proof Sketch]
Pool$_1$ and Pool$_2$ can earn rewards in cases \circled{A}, \circled{B}, and \circled{C} in Figure~\ref{fig:game newbwh}. 
Case \circled{A} represents when an honest miner in one pool
finds an FPoW.
According to case \circled{A}, Pool$_i$ can earn
\begin{equation}
\frac{\alpha_i-f_i}{1-f_1-f_2}.\notag
\end{equation}
Case \circled{B} represents when only one of the two pools finds
an FPoW in the opponent pool
using infiltration mining and submits it to the opponent pool when another miner finds another valid block.
If the FPoW mined by an infiltration miner of Pool$_i$ in the opponent pool is selected as the main chain (with probability $c_i$), the opponent pool can earn the reward
\begin{equation}
c_if_i\frac{1-\alpha_1-\alpha_2}{1-f_i}.\notag
\end{equation}

The final case shows
when infiltration miners of both pools find FPoWs in each of the opponent pool and someone other than the two pools finds another FPoW.
We define $c_i^{'}$ as the probability that the FPoW from 
Pool$_i$'s infiltration mining in the opponent pool
is selected as the main chain among three branches. 
In case \circled{C}, if the infiltration miner of Pool$_1$ first finds an FPoW in the opponent (Pool$_2$) and the FPoW is selected as the main chain, 
Pool$_2$ can earn the reward 
\begin{equation}
c_{1}^{'}f_1\frac{f_2}{1-f_1}\frac{1-\alpha_1-\alpha_2}{1-f_1-f_2}.\notag
\end{equation}
If the infiltration miner of Pool$_1$ finds another FPoW in Pool$_2$ after an infiltration miner of Pool$_2$ finds an FPoW in Pool$_1$, 
Pool$_2$ can earn the reward
\begin{equation}
c_{1}^{'}f_2\frac{f_1}{1-f_2}\frac{1-\alpha_1-\alpha_2}{1-f_1-f_2},\notag
\end{equation}
when the FPoW found from an infiltration miner of Pool$_1$ is selected as the main chain.
Therefore, in case \circled{C}, Pool$_i$ can earn the reward
\begin{equation}
c_{\neg i}^{'}f_1f_2(\frac{1}{1-f_1}+\frac{1}{1-f_2})\frac{1-\alpha_1-\alpha_2}{1-f_1-f_2} \quad (c^{'}_1 + c^{'}_2 \leq 1).\notag
\end{equation}

Lastly, Pool$_i$ can earn the reward 
\begin{equation}
\frac{R_{\neg i}f_i}{\alpha_{\neg i}+f_i}\notag
\end{equation}
through infiltration mining. 
Based on the above rewards for these cases, the rewards $R_1$ of Pool$_1$ and $R_2$ of Pool$_2$ can be
expressed as Eq.~\eqref{eq:game_pool1} and ~\eqref{eq:game_pool2}, respectively.
\end{myproof}

Next, we show that the game has a unique Nash equilibrium, 
and this equilibrium point does not represent honest mining by 
both players since 
a pool can always earn the extra reward by executing the FAW attack
against a compliant pool. 

\begin{theorem}
The game has a unique Nash equilibrium $(f_1, f_2)$, and
this is either a point satisfying
$\frac{\partial R_1}{\partial f_1}=0$, $\frac{\partial R_2}{\partial f_2}=0$
or a point on a borderline satisfying these restricted conditions.
\end{theorem}

\begin{myproof}[Proof Sketch]
To prove the existence of a Nash equilibrium, it 
suffices to show that 
the second partial derivatives of $R_1$ and $R_2$ for
$f_1$ and $f_2$, respectively, are always negative under the following
conditions: 
\begin{align}
&0\leq f_1\leq\alpha_1 \leq1\notag\\
&0\leq f_2\leq\alpha_2 \leq1\notag\\
&\alpha_1+\alpha_2 \leq 1\notag\\
&0\leq c_1, c_2 \leq 1 \notag\\
&0\leq c_1^{'}+c_2^{'}\leq 1. \notag
\end{align}
Therefore, a unique Nash equilibrium point exists since the
functions are strictly concave under these conditions~\cite{rosen1965existence}.

Next, we find the equilibrium point by using \textit{Best-response dynamics}.
Pool$_1$ and Pool$_2$ start with $(f_1,f_2)=(0,0)$ and 
alternately update these values to the most profitable infiltration mining power. 
If we first update Pool$_1$'s infiltration power $f_1^{(1)}$ to maximize $R_1$, 
then Pool$_2$'s infiltration power $f_2^{(1)}$ would be adjusted to maximize $R_2$ according to $f_1^{(1)}$. 
After that, Pool$_1$'s infiltration power $f_1^{(2)}$ again is updated for
maximizing $R_1$ based on $f_2^{(1)}$.
This process repeats continuously. 
When we generalize this for the $k$-th process, $f_1^{(k)}$ and $f_2^{(k)}$ are
represented by 
\begin{equation}
f_1^{(k)}=\arg\max_{0\leq f_1\leq \alpha_1}R^{'}_1 (f_1,f_2^{(k-1)}), \quad 
f_2^{(k)}=\arg\max_{0\leq f_2\leq \alpha_2}R^{'}_1 (f_1^{(k)},f_2), \notag
\end{equation}
respectively.
If $f_1^{(k)}$ and $f_2^{(k)}$ converge as $k$ approaches infinity, the
values will be in a Nash equilibrium.
The Nash
equilibrium $(f_1, f_2)$ is either a point satisfying
$\frac{\partial R_1}{\partial f_1}=0, \frac{\partial R_2}{\partial f_2}=0$
or a point on a borderline of the possible region.
\end{myproof}

\subsection{Quantitative Analysis}
\label{subsec:game-numerical}

\begin{figure}[ht!]
\centering{
\subfloat[]{\hspace{-4mm}
\includegraphics[width=0.25\textwidth]{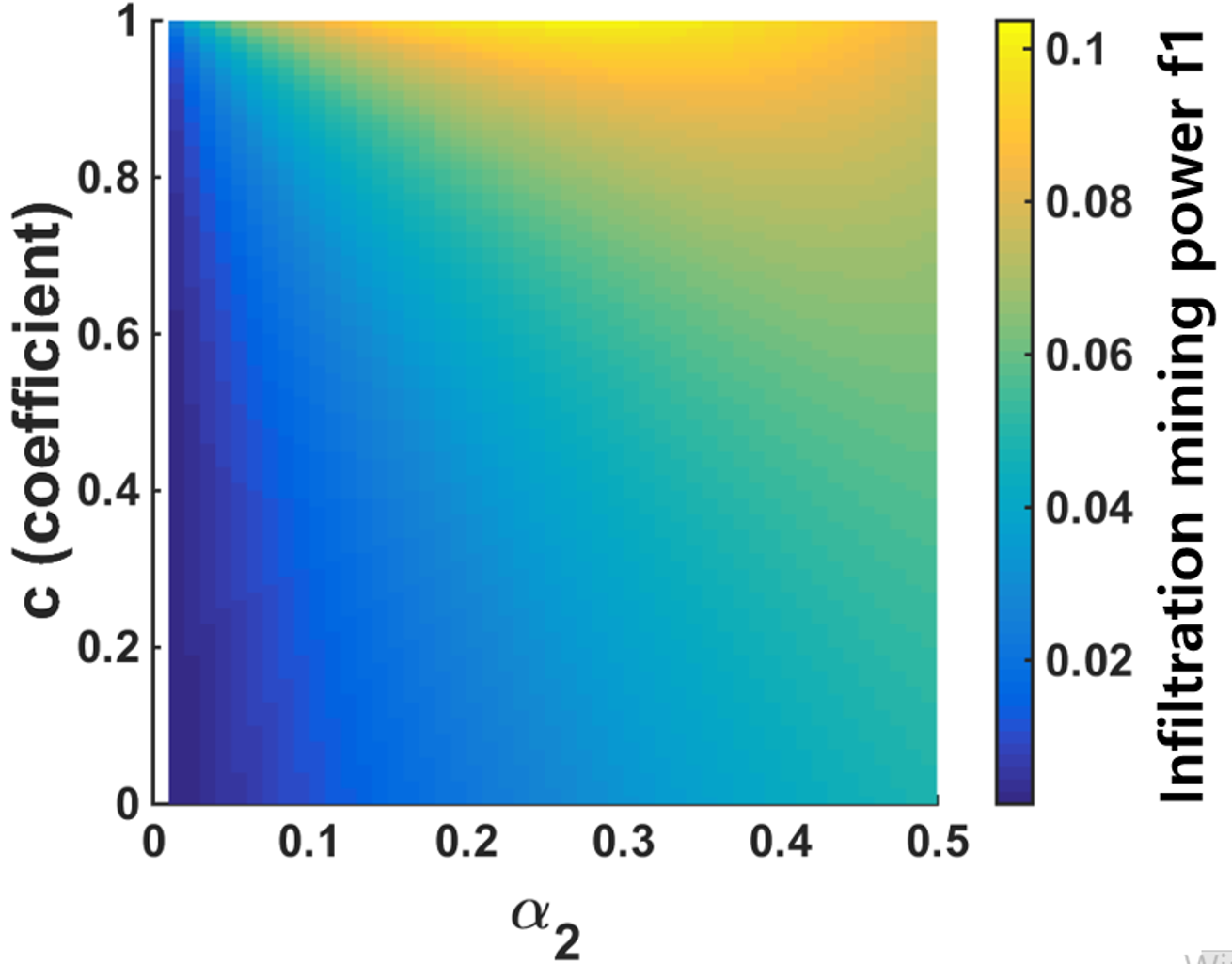}
\label{fig:tau 1_m}
}
\subfloat[]{\hspace{-1mm}
\includegraphics[width=0.25\textwidth]{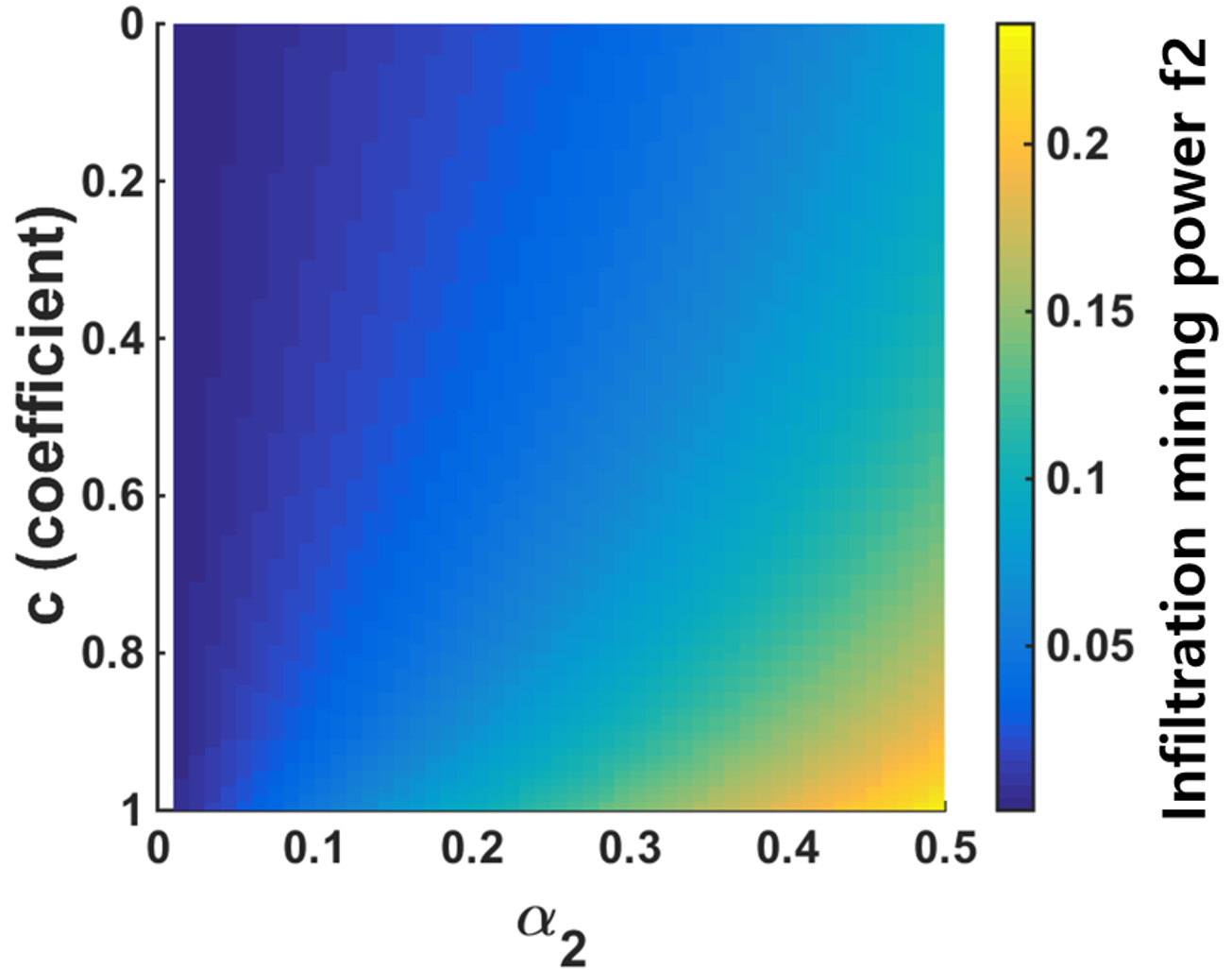}
\label{fig:tau 2_m}
}
}
\centering{
\subfloat[\vspace*{-3mm}]{\hspace{-4mm}
\includegraphics[width=0.25\textwidth]{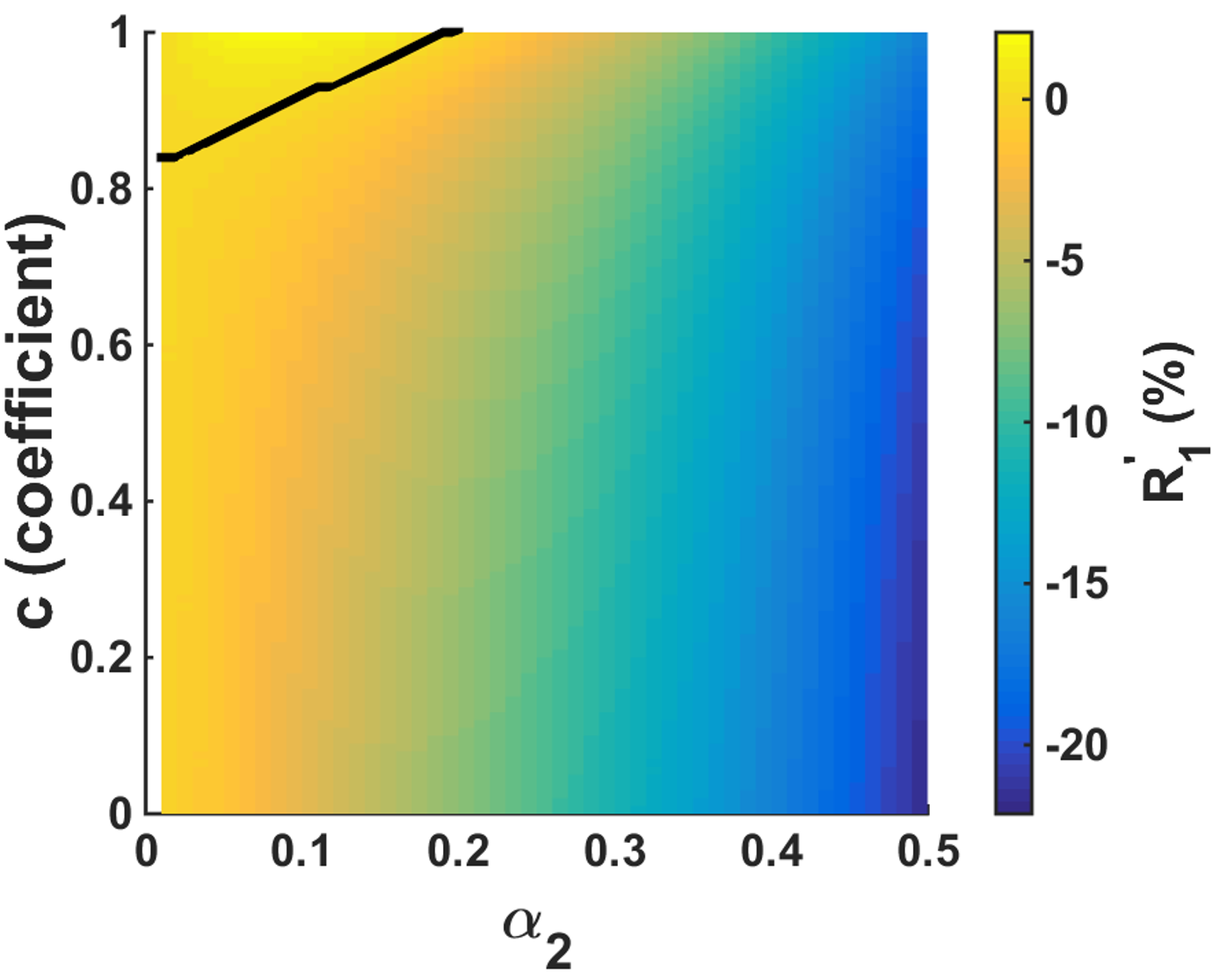}
\label{fig:game result 1}
}
\subfloat[\vspace*{-3mm}]{\hspace{-1mm}
\includegraphics[width=0.25\textwidth]{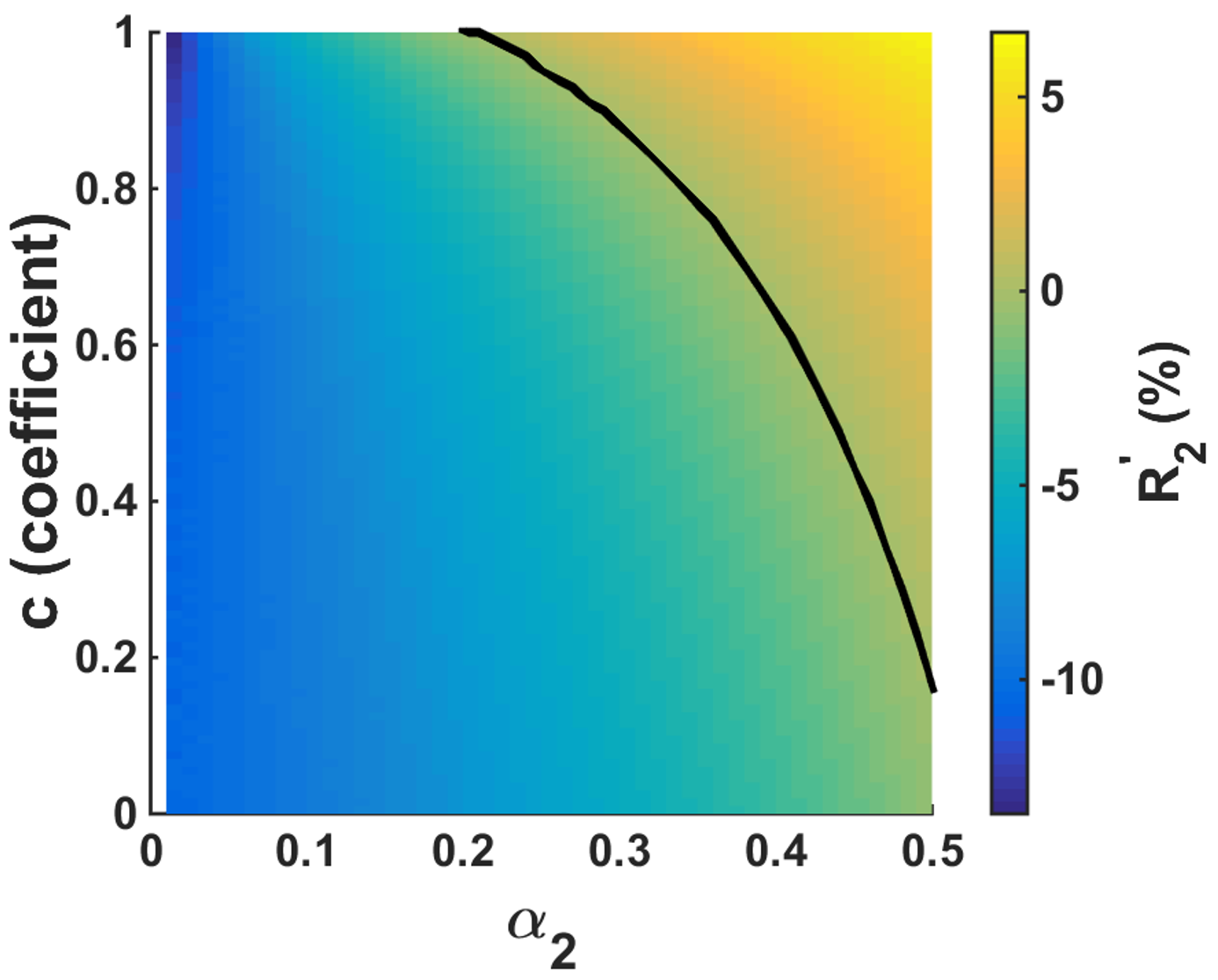}
\label{fig:game result 2}
}
}
\caption{Results of the FAW attack game with varying Pool$_2$'s size $\alpha_2$ and network capability
$c$ where Pool$_1$'s size $\alpha_1$ is 0.2. 
(a) and (b) show the infiltration mining power of 
Pool$_1$ and Pool$_2$ as $f_1$ and $f_2$ in the Nash equilibrium point, respectively.
(c) and (d) represent RERs (\%) $R_1^{'}$ and $R_2^{'}$ for 
Pool$_1$ and Pool$_2$ in the Nash
equilibrium point according to $\alpha_2$ and $c$, respectively. 
Also, the black lines in (c) and (d) are the borderlines at which Pool$_1$ and Pool$_2$ earn the 
same RER as an honest miner, respectively. Above the lines, each pool earns the extra reward, so the prisoner's dilemma does not hold.
\vspace*{-3mm}}
\label{fig:game result}
\end{figure}

We quantitatively analyze the results of the game between two pools in the
Nash equilibrium point. To reduce the parameter dimensions,
we assume that $c_i$ and $c_i^{'}$ are symmetrical for
$i=1,2$ and can be expressed as $c$ and $c/2$, respectively,
while ($0 \leq c \leq 1$). Fig.~\ref{fig:game result}
represents the results of the FAW attack game in terms of $\alpha_2$
and $c$ if $\alpha_1$ is 0.2. Figs.~\ref{fig:tau 1_m}
and~\ref{fig:tau 2_m} show infiltration mining power 
$f_1$ and $f_2$ in the equilibrium. 
Figs.~\ref{fig:game result 1} and \ref{fig:game result 2}
represent RERs (\%) $R_1^{'}$ and $R_2^{'}$ of Pool$_1$ and Pool$_2$
(these parameters are defined as in Section~\ref{subsec:one_quant}) in the
equilibrium, respectively, in terms of $\alpha_2$ and $c$.
The \textbf{black lines} in Figs.~\ref{fig:game result 1} and~\ref{fig:game
result 2} are the borderlines at which Pool$_1$ and Pool$_2$ earn the
same RER as an honest miner, respectively. 
That is, Pool$_1$ and Pool$_2$ can earn the extra reward in the
regions above the black lines in the corresponding figure, while
taking a loss below the black lines.
As a result, Pool$_1$ and Pool$_2$ can win the game if
$(\alpha_2, c)$ is above the black lines in
Figs.~\ref{fig:game result 1} and~\ref{fig:game result 2} when 
Pool$_1$'s size is 0.2. Figs.~\ref{fig:game result 1} and~\ref{fig:game
result 2} also show that the FAW attack game becomes a pool size game, 
because the region above the black line is the case in which 
Pool$_1$'s size is larger than Pool$_2$'s size (and vice versa).


\subsection{Winning Conditions}
\label{subsec:game-cond}

\begin{figure}[ht]
\centering
\includegraphics[width=\columnwidth]{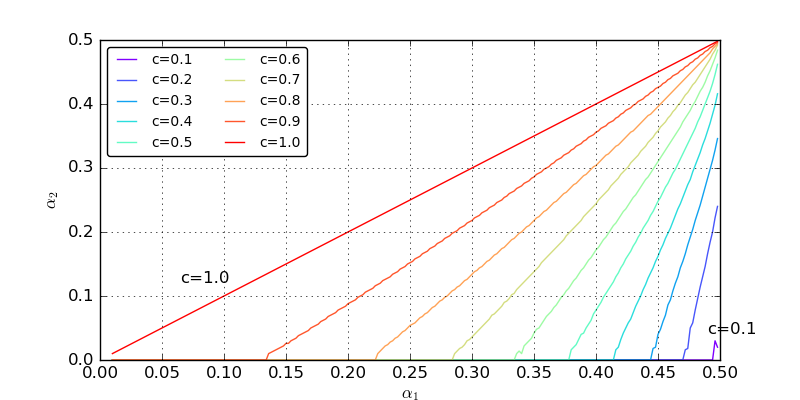}
\caption{Winning conditions for Pool$_1$ with respect to $c$. 
The ten lines represent borderlines at which Pool$_1$ can earn the same 
reward as an honest miner according to $c$. 
The region to the right side of each line represents the winning range of Pool$_1$ in terms of $c$.
Winning conditions for Pool$_2$ are found by swapping the $x$- and $y$-axes.}
\label{fig:win game}
\end{figure}

Eyal discovered that a game between two pools for the BWH attack
brings forth the ``miner's dilemma'', because both suffer a loss in the Nash
equilibrium when their computational power is less than
0.5~\cite{eyal2015miner}. 
In the FAW attack game, the miner's dilemma may not occur, even if 
the size of each of the pools is less than 0.5. 
The region to the right side of each line in
Fig.~\ref{fig:win game} represents the winning range of Pool$_1$
in terms of $c$. The ten lines represent borderlines at which Pool$_1$
can earn the same reward as an honest miner when values of
$c$ vary from 1 to 0.1. When $c$ is 1, the borderline is exactly the
line $\alpha_1=\alpha_2$. In other words, the larger pool
always earns the extra reward, and the smaller pool takes a loss. 
Therefore, the result becomes dependent on pool size,
even in the region where the
miner's dilemma holds in the BWH attack game. Furthermore, the
region in which the miner's dilemma does not hold exists even if $c$ is less than 1. 
In summary, under reasonable conditions for 
two pools' computational power 
and network capabilities, the largest pool earns the extra reward. 
This makes the FAW attack a dominant strategy for any large pool to launch against smaller pools.

\section{FAW Attack vs. Selfish Mining}
\label{Compare}

In this section, we discuss the practicality of the FAW attack in comparison with selfish mining, 
given that both require intentional forks.
Eyal et al.~\cite{eyal2014majority} used the term
$\gamma$ to represent the fraction of the honest network that selects an attacker's block as the main chain in a fork in selfish mining. 
The value of $\gamma$ cannot be 1
because when the intentional fork
occurs, the honest miner who generated a block will select his block,
not that of the selfish miner. Therefore, the value of
$\gamma$ is upper bounded as follows if $\alpha$ is the attacker's
computational power and $o_i$ is the computational power of the honest
node $i$: 
\begin{equation}
\begin{small}
\gamma \le 1 - \sum_{i}{\frac{o_i}{(1-\alpha)}\left(1-\frac{o_i}{(1-\alpha)}\right)} \le1 - \sum_{i}{\frac{o_i^2}{(1-\alpha)^2}}<1 - \sum_{i}o_i^2.\notag
\end{small}
\end{equation}
Note that the total power of honest nodes is $1-\alpha$ 
(i.e., $\sum_{i}o_i=1-\alpha$). 
Therefore, if a selfish miner belongs to the Unknown group in
Table~\ref{tab:pool} (i.e., is a solo miner or a closed pool), the value
of $\gamma$ is loosely upper bounded by 0.89 according to 
Table~\ref{tab:pool}.
Eyal et al.~\cite{eyal2014majority} stated that an attacker needs at least
$\frac{1-\gamma}{3-2\gamma}$ computational power for selfish mining
to be profitable. 
As a result, the attacker needs computing power of at least
0.09 even when her network capability is
optimal. However, this power is too high for most solo miners or closed pools. 
For them,
selfish mining is not profitable.
In contrast, the FAW attack is
always profitable regardless of an attacker's computational power
(see Sections~\ref{Onepool} and~\ref{multipool}). This makes the FAW attack
more practical for a solo miner or a closed pool.

Next, we consider a case in which a selfish miner is an open pool manager.  
Here, the cost for
selfish mining may not be very high for the attacker.  
However, the
selfish open pool manager must be concerned about 
whether honest miners will leave
her pool by disclosing direct evidence before she earns the extra reward, 
because honest miners do not want to destabilize Bitcoin.
Indeed, honest miners belonging to the attacker's pool can
easily detect that their pool manager is a selfish mining attacker in two
ways. First, if the manager does not propagate blocks immediately
when honest miners generate FPoWs, the honest miners will 
know that their pool manager is an attacker. 
Second, the blockchain has an abnormal shape when a selfish miner
exists; Bitcoin miners can determine 
which open pool has caused the abnormal shape 
because which open pool has found each block is public information. 
This information is provided by several services~\cite{blockchain, hashdist}.
For example, when one branch of a fork contains consecutive blocks
generated by the attacker's pool in a short time period, 
the pool may be suspect. 
Even if the attacker tweaks her strategy to evade detection 
by releasing her blocks gradually, one branch of the fork will still contain 
consecutive blocks generated by the attacker's pool. 
Therefore, all participants in Bitcoin including honest miners in the pool can
detect that the pool is a selfish miner before she earns the extra reward. 
As a result, open pool
managers are unlikely to execute selfish mining.  

When the FAW attack occurs and the attacker is an open pool manager, the fork rate may increase;
therefore, detecting the existence of the FAW attack may not be
difficult. However, identifying the attacker is more challenging
than with selfish mining 
because if an honest miner in her pool generates an FPoW, 
the FAW attacker propagates the block
immediately, which differs from selfish mining.  
In addition, since the infiltration miner in the target pool generates forks intentionally by propagating
FPoWs to the target pool, the identity of the target but not the
attacker is disclosed. 
In other words, the attacker's pool looks innocent, and 
meanwhile the target pool looks strange due to 
its high rate of forks.

\section{Network Capability $c$}
\label{Practicality}

For an attacker to execute an FAW attack, 
she needs to know
some information in advance.  First, an attacker's optimal $\overline{\tau}$ depends not only on
the attacker's computation power, but also on that of the target pool.
Therefore, she must know the target pool's computational power. 
Its approximate value can be
obtained from the current computational power distribution~\cite{hashdist}, which is public information.

However, she also needs to know the value of network capability $c$ in order to adopt an optimal
$\overline{\tau}$ in Eq.~\eqref{eq:op_tau}.  
The term $c$ is the
probability that an attacker's FPoW from infiltration mining will be
selected as the main chain. In this section, a possible range of $c$ is first given, 
and then attacker behavior for a
constant yet unknown $c$ is discussed. We extend this discussion to the 
case in which $c$
changes frequently. The results are promising. We show that
the FAW attack still improves upon the BWH attack, even if $c$ is unknown.
Furthermore, interestingly, the range to which 
the miner's dilemma applies decreases 
compared to when $c$ is known in the FAW attack game.

\vspace*{2mm}
\noindent\textbf{The Possible Range of $c$: }
The value of network capability $c$ is greater than or equal to 0 by definition. 
\textit{In practice, the value of $c$ is positive, because it is possible for 
an attacker to listen to external block propagation
faster than the manager using Sybil nodes.}
Moreover, if the target pool's manager behaves rationally, 
the minimum value of $c$ (in Section~\ref{Onepool}) is the sum of the computational power of the attacker 
and the target pool because the attacker and target pool select her FPoW 
found through infiltration mining. 
Here, the manager's rational behavior is to select the block found by 
infiltration miner in his pool as the main chain 
even if the infiltration miner propagates this FPoW to the manager 
\textit{right after} he notices that an external miner has found a block.
In the same manner, since two players in the FAW attack game 
are rational, the value of $c$ in the FAW attack game between 
two pools is lower bounded by $\alpha_1+\alpha_2$.
The maximum value of $c$ also depends on computational power
distribution in the Bitcoin network because an honest miner
(neither belonging to the target pool nor representing the attacker) who generates
an FPoW selects his own block, not the block from the attacker's FPoW
from infiltration mining.  
Therefore, even if an attacker has
optimal network capability, the maximum value of $c$ in Sections~\ref{Onepool}
and~\ref{multipool} is upper bounded by \vspace{-3mm}
\begin{equation}
c=\sum_{j}\frac{o_j}{1-\alpha-\beta}(1-o_j)=1-\frac{\sum_{j} o_j^2}{1-\alpha-\beta}\notag
\end{equation}
when $o_j$ is the computational power of an external honest miner node $j$.
Note that the total computational power of honest miners $\sum_{j} o_j$
is $1-\alpha-\beta$.  Also, the value of
$c$ in Section~\ref{Game} is upper bounded by \vspace{-3mm}
\begin{equation}
c=\sum_{j}\frac{o_j}{1-\sum_{i=1\sim n}\alpha_i}(1-o_j)=1-\frac{\sum o_j^2}{1-\sum_{i=1\sim n}\alpha_i}\notag
\end{equation}
when the game participants are $n$ open pools. 
In this case, this condition $\sum_{j} o_j=1-\sum_{i=1\sim n}\alpha_i$ is 
satisfied. 

For example, if two pools, F2Pool and BitFury, with computational
powers of $20\%$ and $10\%$, respectively, as in Table~\ref{tab:pool},
participate in the FAW attack game, the maximum value of $c$ is about
0.914. Note that this case does not fall into the miner's dilemma,
and, therefore, the game becomes the pool size game.  Moreover, when
the power of honest miners ($o_j$) is evenly distributed among many nodes,
$c$ may be closer to 1. Thus, if an attacker executes the FAW
attack against all open pools, or if all
open pools participate in the FAW attack game, 
the maximum value of $c$ may be close to 1. 

In addition, network capability $c$ can be expressed as 
$\gamma(1-\alpha-\beta)+\alpha+\beta$ when the target is one pool, and the target manager 
behaves rationally in order to reduce loss 
(applying the network capability 
term $\gamma$ used in prior research~\cite{eyal2014majority, nayak2016stubborn}). 

\vspace*{2mm}
\noindent\textbf{Constant $c$: }
\label{subsubsec:const}
We first assume that the value of $c$ is constant but unknown to the FAW
attacker against one pool.
Under such conditions, she cannot apply Eq.~(\ref{eq:op_tau}) directly
because optimal $\overline{\tau}$ depends on the value of $c$.  
However, she
knows that the value of $c$ is greater than or equal to 0 if the target pools' 
managers are honest. Thus, she
can choose $\overline{\tau_{0}}$, obtained from
Eq.~(\ref{eq:op_tau}) substituting $c$ with 0.  
In such a case,
the attacker can still earn a greater reward than the BWH attacker.
The FAW attacker's reward $R_a(\overline{\tau_0})$ is 
\begin{equation}
\max_{\tau}(R_{BWH})+c\overline{\tau_0}\alpha\cdot
\frac{1-\alpha-\beta}{1-\overline{\tau_0}\alpha}\cdot\frac{\overline{\tau_0}\alpha}{\beta+
\overline{\tau_0}\alpha}, \notag
\end{equation}
which is lower bounded by the BWH attacker's reward $R_{BWH}$. 

If the target pool's manager is rational, the attacker repeats the above process, substituting $c$ with $\alpha+\beta$, the minimum value of $c$ 
in Eq.~(\ref{eq:op_tau}). Thus, she uses 
$\overline{\tau_{\alpha+\beta}}$ as the value of $\tau$.
Then, the FAW attacker earns extra reward that is
certainly more than that for the BWH attacker. 
Note that the attacker can test whether the manager 
is rational by submitting a stale FPoW. 
The attacker can also learn about $c$, investigating
the relationship between long-term and theoretical rewards for the
minimum value of $c$, when we assume that $c$ is constant.  
As a result, she can find an
optimal $\overline{\tau}$ (Eq.~\eqref{eq:op_tau}), and her reward
converges to the maximum value of $R_a$.  

\begin{figure}[ht]
\centering
\includegraphics[width=\columnwidth]{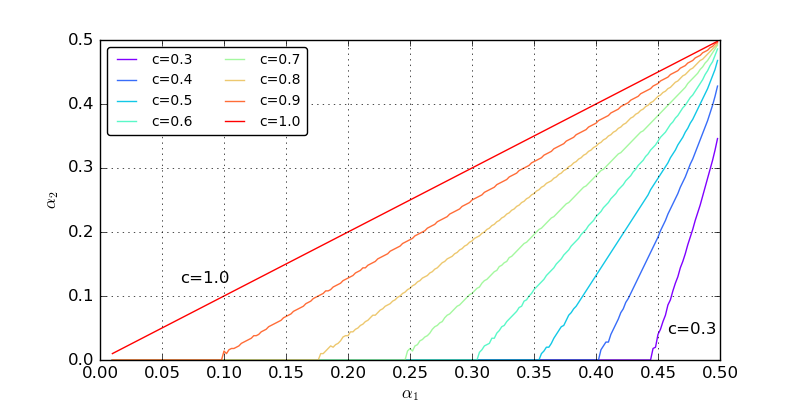}
\caption{ The winning condition of Pool$_1$ versus $c$.
Ten lines represent borderlines at which Pool$_1$ can earn the same 
reward as an honest miner according to $c$. 
The region to the right side of each line represents the winning range of Pool$_1$ in terms of $c$.
Pool$_2$'s winning conditions are found by swapping the $x$- and $y$-axes.
\vspace*{-2mm}}
\label{fig:win game2}
\end{figure}

\vspace*{2mm}
\noindent\textbf{Frequently Changing $c$: }
\label{subsubsec:unstable}
The Bitcoin network often changes, with the power
distribution and number of nodes shifting as well~\cite{hashdist,bitnodes}. 
Thus, the value of $c$ may also change. When an attacker
executes the FAW attack against one pool and the pool manager is honest,
as in the above case, she must 
use $\overline{\tau_{0}}$ as the value of $\tau$. In fact, the
attacker may ignore the fact that $c$ changes. For example, she
may assume $c=0$ and choose an optimal strategy. Applying this
strategy to the FAW attack against the four open pools in
Table~\ref{tab:pool}, she can earn an RER
of up to 3.99\%. Therefore, the FAW attack improves her RER by
up to 34.62\% of that for the BWH attack even if the attacker knows
nothing about $c$.
Moreover, in the FAW attack game between two pools, two pools may assume
$c=\alpha_1+\alpha_2$, which is the minimum value of $c$, in practice.
Using the FAW game between F2Pool (Pool$_1$) and BTCC Pool (Pool$_2$)
in Table~\ref{tab:pool} as an example, 
both managers may assume $c=0.3$. 
Then, the
winning conditions for F2Pool (Pool$_1$) are shown in Fig.~\ref{fig:win
  game2}. Furthermore, compared to Fig.~\ref{fig:win game}, Fig.~\ref{fig:win
  game2} shows how the region affected by the miner's dilemma
decreases. Indeed, when the assumed value of $c$ decreases, 
the region affected by the miner's dilemma decreases as well.

\vspace*{-1mm}
\section{Discussion}
\label{Discussion}

\subsection{Rational Manager}
\label{subsec:more}

In the FAW attack, an attacker submits an FPoW to a manager to
generate a fork when an external miner broadcasts a block. 
For her block to be selected, she must quickly notice
the external block propagation using Sybil nodes. If she detects the
propagation 
before the pool manager, a
fork can be caused naturally, from the manager's perspective. 
When she learns of the propagation from the manager 
(instead of detecting it first),
she submits her FPoW immediately. In this case, an honest
manager regards the attacker's FPoW as stale and invalidates it because he
knows a new round has already started. However, a
rational manager may not act in accordance with the protocol, 
since \textit{it would always
be beneficial for him to submit a local FPoW}.
We already proved that the manager's behavior can decrease 
his pool's loss, as in Section~\ref{Onepool}. 
This behavior decreases the manager's loss and increases
the attacker's reward as a side-effect. Note that in the FAW attack game
in Section~\ref{Game}, since two pools are attacking each other,
both managers are rational.
Therefore, they always propagate a block found by 
the opponent's infiltration miner in their own pool, 
even if they received a block from an external 
miner first.

\subsection{Detecting FAW Attacks and Attackers}
\label{Detection}

We showed that FAW attacks provide greater rewards to 
attackers than existing BWH attacks. 
From the target pool's perspective, detecting
infiltration mining and identifying the attacker are important. 
Indeed, the FAW attack is easier to detect than the BWH attack because of 
the high fork rate.
Additionally, the manager should suspect 
and expel any miner who submits stale
FPoWs, rather than paying out the reward for the current round.
Note that rewards for previous rounds cannot be returned to the manager 
because of the properties of Bitcoin. 
The attacker may easily
launch the attack using many Sybil nodes with many churns, 
replacing the expelled miner. 
This strategy allows the attacker to receive rewards without 
being greatly affected by the manager behavior, 
even if her FAW attack is detected and her infiltration miner is expelled. 
For example, assuming that an attacker infiltrates a target pool with 
$L$ infiltration miners, each with different worker ID and password,
if the $L$-th infiltration miner 
is detected by the manager, the remaining $L-1$ miners can still earn rewards.
Then the attacker's reward is lower bounded by
\begin{small}
\begin{equation}
\frac{(1-\tau)\alpha}{1-\tau\alpha}
+\frac{\beta}{1-\tau\alpha}\cdot\frac{(L-d)\tau\alpha}{L\beta+(L-d)\tau\alpha}+c\tau\alpha\cdot\frac{1-\alpha-\beta}{1-\tau\alpha}\cdot\frac{(L-d-1)\tau\alpha}{L\beta+\tau\alpha(L-d-1)}.\notag 
\end{equation}
\end{small}
Here $d$ is defined as the average number of FPoWs, which are submitted by
infiltration miners for a while untill the pool earns the reward for one block, and but not selected as the main chain. The value of $d$ can be expressed as 
 \begin{equation}
\frac{(1-c)\gamma\alpha (1-\alpha-\beta)}{\beta+c \gamma\alpha (1-\alpha-\beta)}. \notag
\end{equation}

Therefore, the more infiltration miners are used (i.e., the more $L$ increases), 
the less detection affects the attacker. 
She may continue the FAW attack by substituting the $L$-th miner 
with another infiltration miner.
Thus, the FAW attacker's reward is still better than
the BWH attacker's for a properly chosen $L$ because 
the minimum value of $c$ is positive in practice.
Additionally, an attacker can twist the FAW attack by propagating 
the withheld FPoW only when she notices external block propagation 
faster than the manager if the manager is honest.
Also, she can hide her IP address by using hidden services such as Tor.

\vspace*{-1mm}

\subsection{Countermeasures}
\label{Countermeasure}

Even if we focus on the FAW attack against Bitcoin, 
other proof-of-work cryptocurrencies such as Ethereum~\cite{wood2014ethereum}, Litecoin~\cite{litecoin}, Dogecoin~\cite{dogecoin}, and Permacoin~\cite{miller2014permacoin} are also vulnerable to the FAW attack.
Especially, Ethereum adopts a protocol based on GHOST~\cite{sompolinsky2015secure} unlike Bitcoin.
Therefore, the FAW attacker's reward in the case of Ethereum should be
recalculated.
Because the FAW attack breaks the dilemma and is more practical than selfish
mining, it can be launched from large pools in these cryptocurrencies. 

We discuss possible countermeasures against the FAW attack. 
First, an approach must satisfy \textit{backward compatibility} 
in order to be a practical defense mechanism.
Backward compatibility means miners who have not upgraded their
mining hardware can still mine after the measures are 
implemented~\cite{zhang2017publish}, 
retaining miners' current mining hardware investments~\cite{twophase}. 
This is important because Bitcoin's security is directly
related to total mining power.
Therefore, it is impractical to make a major change to the Bitcoin protocol for defense.
The \textit{two-phase PoW} protocol, called \textit{Oblivious Shares}, presented by Rosenfeld ~\cite{rosenfeld2011analysis}
which can defend against both BWH and FAW attacks
is impractical on these grounds. 

Second, to prevent FAW attacks, it is not sufficient to just detect 
the infiltration miner. As described in Section~\ref{Detection}, 
detection rarely affects 
the FAW attacker. For detection, one may consider the following
mechanism: 

\textit{``Mining pool managers could provide a \textit{beacon} value that is updated very frequently (i.e., every couple of seconds) and only give points for PPoWs that include a recent \textit{beacon} value.''} 

This defense has an effect only when an attacker notices external block
propagation faster than 
the manager, subsequently propagating a withheld FPoW.
(If the attacker notices the propagation after the manager, 
the manager already knows that the FPoW is stale.)
In this case, the manager may notice the FPoW is stale 
because it includes 
a stale \textit{beacon} value. However, the manager would still propagate
a valid block based on the FPoW. 
Note that this credible behavior does not deviate from Bitcoin protocol 
because the manager received the internal FPoW before the external one.
Then, as mentioned in Section~\ref{Detection}, the remaining infiltration miners  (e.g., $L-1$ infiltration miners in Section~\ref{Detection}) receive a reward 
even if the infiltration miner (e.g., the $L$-th infiltration miner), 
who submitted the FPoW, is expelled.
As a result, the attacker still earns a higher reward than the BWH attacker.

Another \textit{two-phase PoW}~\cite{twophase} proposed by Eyal and Sirer can be used to defend against FAW attacks. 
This defense has better backward compatibility than Rosenfeld's \textit{Oblivious Shares}~\cite{rosenfeld2011analysis}. 
In both schemes, a miner does not know whether his PPoW is a 
valid block because generating a PoW is divided into two steps.
However, the Bitcoin community would not like to adopt the two-phase PoW 
proposed by Eyal and Sirer as well~\cite{eyal2015miner}. 
Such an approach would be inconvenient for closed pools and solo miners 
who 
are not concerned about being targets of BWH and FAW attacks.
For pool managers, this protocol increases the cost of pool operation.
Moreover, pool miners are concerned about \textit{block withholding} by 
pool managers. 
A rational manager can waste miners' power by withholding blocks in 
her pool and then earn higher rewards through solo mining. 
If the malicious manager throws away all blocks found by miners, 
miners can detect it in a short time period. However, when the manager throws 
away just a part of the blocks (e.g., 5\%), miners cannot detect it for a long
time. 
Such behavior can be seen as a new variant of the BWH attack. 
As a result, \textit{two-phase PoW} proposed by Eyal and Sirer is hardly suitable for adoption by the Bitcoin system.
Note that \textit{Oblivious Shares} also has drawbacks described above.

Eyal~\cite{eyal2015miner} and Luu et al.~\cite{luu2015power} 
have introduced several
countermeasures against BWH attacks.
A joining fee was one such measure, 
but Eyal concluded that miners prefer flexibility.
A honeypot trap was also proposed, 
but the idea was quickly dropped due to high overhead. 
Moreover, even if this idea is practical, 
BWH and FAW attacks can still be profitable
if an attacker uses many ($L$) infiltration miners. 
As established in Section~\ref{Detection}, the remaining $L-1$ miners 
can still receive rewards even if the $L$-th miner is detected.
Indeed, the reward for a BWH attacker given the honeypot trap 
is lower bounded by 
\begin{equation}
\frac{(1-\tau)\alpha}{1-\tau\alpha}
+\frac{\beta}{1-\tau\alpha}\cdot\frac{(L-d)\tau\alpha}{L\beta+(L-d)\tau\alpha}\,
\text{ if } d=\frac{\gamma\alpha (1-\gamma\alpha)}{\beta}.\notag
\end{equation} 
Both studies also proposed new reward systems to incentivize
miners to submit FPoWs immediately. To prevent FAW attacks, 
we may consider a new reward system. 
A pool
miner who finds an FPoW (as opposed to a PPoW) can receive a bonus from the
manager. If, for example, the manager receives 1~BTC for
each block, the miner who finds an FPoW may receive 0.1~BTC, with 
0.9~BTC distributed among all miners in proportion to their work shares.
Theorem~\ref{thm reward} shows this defensive reward scheme against 
FAW attacks. 

\begin{theorem}
\label{thm reward}
If a reward fraction $t$ of the total reward (e.g., 1 BTC)
for one valid block is given to the miner who finds an FPoW, then
the attacker's reward, $R_a$, is 
\begin{equation}
\label{eq:reward_t}
\resizebox{0.91\hsize}{!}{%
  $
\frac{(1-\tau)\alpha}{1-\tau\alpha} + \frac{\beta}{1-\tau\alpha}\cdot (1-t)
\cdot\frac{\tau\alpha}{\beta+\tau\alpha}
+ c\tau\alpha\cdot \frac{1-\alpha-\beta}
{1-\tau\alpha}\cdot(t+(1-t)\frac{\tau\alpha}{\beta+\tau\beta}). 
$}
\end{equation}
When the manager chooses 
\begin{equation}
t \geq \frac{1}{2(1-c_{max}(1-P))}\notag
\end{equation}
for the pool's current computational power, $P$, $R_a$ is always 
less than $\alpha$.
\end{theorem}

\begin{myproof}[Proof Sketch]
The attacker can still earn the reward $\frac{(1-\tau)\alpha}{1-\tau\alpha}$
through innocent mining. 
When an honest miner finds an FPoW in the target pool, 
she gets paid a fraction of the reward $1-t$ according to her infiltration mining power.
Because the probability that an honest miner finds an FPoW in 
the target pool is $\frac{\beta}{1-\tau\alpha}$, the attacker's reward from 
the case is 
\begin{small}
\begin{equation}
\frac{\beta}{1-\tau\alpha}\cdot (1-t)
\cdot\frac{\tau\alpha}{\beta+\tau\alpha}.\notag
\end{equation}
\end{small}
Next, if she submits an FPoW in order to generate a fork, 
she can receive the reward including $t$. 
Therefore, the attacker's reward for the case is 
\begin{small}
\begin{equation}
c\tau\alpha\cdot \frac{1-\alpha-\beta}
{1-\tau\alpha}\cdot(t+(1-t)\frac{\tau\alpha}{\beta+\tau\beta}).\notag
\end{equation}
\end{small}
Considering above all cases, the total reward $R_a$ for the attacker is Eq.~\eqref{eq:reward_t}.

Then we find the condition for $t$ which makes $R_a$ less than the
reward $R_h$ of an honest miner, who possesses the computational power, $\alpha$. 
\begin{equation}
\resizebox{0.91\hsize}{!}{%
  $
\frac{(1-\tau)\alpha}{1-\tau\alpha} + \frac{\beta}{1-\tau\alpha}\cdot (1-t)
\cdot\frac{\tau\alpha}{\beta+\tau\alpha}
+ c\tau\alpha\cdot \frac{1-\alpha-\beta}
{1-\tau\alpha}\cdot(t+(1-t)\frac{\tau\alpha}{\beta+\tau\beta})<\alpha\notag
$}
\end{equation}
\begin{align}
&\Leftrightarrow\frac{\beta+\tau\alpha-\tau^2\alpha+c\tau^2\alpha(1-\alpha-\beta)}
{(1-\tau\alpha)(\beta+\tau\alpha)} -t \frac{\beta\tau-c\tau\beta (1-\alpha-\beta)}{(1-\tau\alpha)(\beta+\tau\alpha)}<1 \notag\\
&\Leftrightarrow -\tau^2\alpha + c\tau^2 \alpha (1-\alpha-\beta)+\tau\alpha\beta+
\tau^2\alpha^2<t\beta\tau(1-c(1-\alpha-\beta))\notag\\
&\Leftrightarrow
\frac{\tau\alpha(c(1-\alpha-\beta)-1)+\tau\alpha^2+\alpha\beta}{\beta(1-c(1-\alpha-\beta))} < t \label{eq:tcon}
\end{align}
Therefore, for $R_a$ to be less than $\alpha$, 
$t$ has to satisfy Eq.~\eqref{eq:tcon} for all possible values of $\tau$ and $c$.
(Note that the range of $\tau$ is between 0 and 1, and $c$
ranges from 0 to $c_{max}$.) 
In other words, $t$ has to be greater than 
the maximum of the left-hand side of Eq.~\eqref{eq:tcon} for $\tau$ and $c$.
The maximum can be derived as follows. 
\begin{align}
& \frac{\tau\alpha(c(1-\alpha-\beta)-1)+\tau\alpha^2+\alpha\beta}{\beta(1-c(1-\alpha-\beta))}\notag\\
& = \tau\alpha\left(\frac{\alpha(1-c)+c(1-\beta)-1}{\beta(1-c(1-\alpha-\beta))}\right)+\frac{\alpha}{1-c(1-\alpha-\beta)}\notag\\
&\le \frac{\alpha}{1-c(1-\alpha-\beta)} \,\,\,(\because \alpha(1-c)+c(1-\beta) \leq \frac{1}{2}(1-c)+c\leq 1)\notag\\
&\le\frac{\alpha}{1-c_{max}(1-\alpha-\beta)}\notag
\end{align}
Thus, the condition of $t$ needed to prevent the FAW attack are 
\begin{equation}
  \frac{\alpha}{1-c_{max}(1-\alpha-\beta)}\leq t. 
\label{eq:tcon2}
\end{equation}
The left-hand side of Eq.~\eqref{eq:tcon2} is the same as the computational power $\alpha$ of an
attacker when $c_{max}$ is zero. 
This particular case is equivalent to a defensive reward
system for the BWH attack proposed by Luu et al.~\cite{luu2015power}.

Indeed, because the manager does not know who the attacker is, he does not
know either $\alpha$ or $\beta$. However, he can know 
$\beta+\tau\alpha$ as his pool's current computational power. 
Thus, we express the condition of $t$ as an equation related to the current
pool's computational power.
When the pool's current computational power is $P$, 
the left-hand side of Eq.~\eqref{eq:tcon2} is upper bounded by 
\begin{equation}
\frac{\alpha}{1-c_{max}(1-P)}.\notag
\end{equation} 
Because $\alpha$ is less than 0.5,
the value is less than 
\begin{equation}
\label{eq:t_max}
\frac{1}{2(1-c_{max}(1-P))}. 
\end{equation}
As a result, if $t$ is greater than Eq.~\eqref{eq:t_max}, 
$R_a$ is less than $\alpha$.
\end{myproof}

This theorem shows that the manager can make honest mining 
more profitable than the FAW attack by choosing $t$ properly.
Unfortunately, miners may hesitate to join pools using this
reward system because of the high reward variance. 
We may also consider a reward system in which pool miners get 
a wage for multiple rounds once. Damage to the attacker 
due to detection would be more visible even if the damage decreases as 
the number of infiltration miners (i.e., $L$) increases.
However, this scheme also causes high reward variance,
which might make it difficult for
the pool manager to attract more power. 
Therefore, he should be cautious about adopting this new
reward system, even if it can decrease the risk of the FAW attack.

\section{Conclusion}
\label{Conclusion}

In this paper, we have proposed FAW attacks in which
an attacker withholds a block in a target pool and submits
it when an external miner propagates a valid block.
Such an attack can generate an intentional fork.
Our attack not only improves the practicality of selfish mining
but also yields rewards equal to or greater than those of BWH
attacks.
Unlike the ``miner's dilemma'' that arises in a BWH attack game, 
an FAW attack game can produce a clear winner in the Nash equilibrium point
-- the larger mining pool gains while the smaller pool loses.
Interestingly, rational behavior of the target pool manager also makes
FAW attacks more profitable.
Participants in the Bitcoin network want a cheap and efficient
defense against attacks, including FAW attacks, without introducing major
changes to the Bitcoin protocol or causing side-effects.
Unfortunately, we cannot find such a defense, and 
discovering a solution remains an open problem.
Therefore, we leave it as a future work.
The irrelevance of the miner's dilemma unlike BWH attacks 
and practicality unlike selfish mining
means that proof-of-work cryptocurrencies are expected to see 
large miners executing FAW attacks.

\section*{Acknowledgement}
This research was supported by the MSIT (Ministry of Science and ICT), Korea, under the ITRC (Information Technology Research Center) support program (IITP-2017-2015-0-00403) supervised by the IITP (Institute for Information \& communications Technology Promotion).

\newpage
\bibliographystyle{ACM-Reference-Format}
\bibliography{references}

\begin{appendices}
\section{}
\label{sec:algorithm}

\begin{algorithm}[h]
\captionof{algorithm}{FAW attack against one pool\label{FAWone}}
\begin{algorithmic}[1]
  \State $A$: The miner set of an attacker
  \State $P$: The miner set of a target pool
  \State $F_k$: The $k$-th found FPoW for one round
  \State $X \leftarrow \texttt{work}(Y)$: The miner set $Y$ finds FPoW $X$
  \State $Y \leftarrow \texttt{submit}(X)$: FPoW X is submitted to the manager of $Y$
  \State $\texttt{publish}(Y, X)$: The manager of $Y$ publishes FPoW $X$
  \State $\texttt{discard}(X)$: An attacker discards FPoW $X$
  \medskip
  \Function{round}{}
    \State $k=1$
    \BState \emph{Generate a Fork}:
    \If {$F_k \leftarrow \texttt{work}(A\cap P^c)$}
      \State $\texttt{publish}(A,F_k)$ \Comment{Case A}
    \ElsIf {$F_k \leftarrow \texttt{work}(A^c\cap P)$}
      \State $P \leftarrow \texttt{submit}(F_k)$
      \State $\texttt{publish}(P,F_k)$ \Comment{Case B}
    \ElsIf {$F_k \leftarrow \texttt{work}(A^c\cap P^c)$}
    \If {$k\ne 1$}
      \State $\texttt{publish}(A^c\cap P^c,F_k)$
      \State $P \leftarrow \texttt{submit}(F_1)$
      \State $\texttt{publish}(P,F_1)$ \Comment{Fork, Case C}
    \Else 
      \State $\texttt{publish}(A^c\cap P^c,F_k)$ \Comment{Case D}
    \EndIf
    \Else 
        \State $F_k \leftarrow \texttt{work}(A\cap P)$
	\If {$k\ne1$}
	    \State $\texttt{discard}(F_k)$
	\EndIf
      \State $k$++
      \State goto \emph{Generate a Fork}
    \EndIf
  \EndFunction
\end{algorithmic}
\end{algorithm}

\if false

\vspace{15pt}

\begin{algorithm}[t]
\captionof{algorithm}{FAW attack against one pool\label{FAWone}}
\begin{algorithmic}[1]
  \State $A$: The miner set of an attacker
  \State $P$: The miner set of a target pool
  \State $F_k$: The $k$-th found FPoW
  \medskip
  \Function{oneround}{}
    \If {$F_1 \leftarrow \texttt{work}(A\bigcap P^c)$}
        \State $\texttt{publish}(A,F_1)$ \Comment{Case A}
      \ElsIf {$F_1 \leftarrow \texttt{work}(A^c\bigcap P)$}
        \State $\texttt{publish}(P,F_1)$ \Comment{Case B}
      \ElsIf {$F_1 \leftarrow \texttt{work}(A^c\bigcap P^c)$}
        \State $\texttt{publish}(A^c\bigcap P^c,F_1)$ \Comment{Case D}
      \Else {$F_1 \leftarrow \texttt{work}(A\bigcap P)$}
        \BState \emph{Generate a Fork}:
      \State $k$++
      \If {$F_k \leftarrow \texttt{work}(A\bigcap P^c)$}
        \State $\texttt{publish}(A,F_k)$ \Comment{Case A}
      \ElsIf {$F_k \leftarrow \texttt{work}(A^c\bigcap P)$}
        \State $\texttt{discard}(F_1)$
        \State $\texttt{publish}(P,F_k)$ \Comment{Case B}
      \ElsIf {$F_k \leftarrow \texttt{work}(A^c\bigcap P^c)$}
        \State $\texttt{publish}(A^c\bigcap P^c,F_k)$
        \State $\texttt{publish}(P,F_1)$ \Comment{Case C}
      \Else {$F_k \leftarrow \texttt{work}(A\bigcap P)$}
        \State $\texttt{discard}(F_k)$
        \State goto \emph{Generate a Fork}
      \EndIf
    \EndIf
  \EndFunction
\end{algorithmic}
\end{algorithm}
\fi

\newpage

\begin{algorithm}[h]
\captionof{algorithm}{FAW attack against $n$ pools\label{FAWmultiple}}
\begin{algorithmic}[1]
  \State $A$: The miner set of an attacker
    \State $P_j$: The miner set of a target pool $j$
  \State $P$: $\cup P_j$
  \State $F_k$: The $k$-th found FPoW for one round
  \State $F_{wh, i}$: The FPoW found by $A$ in the pool $i$
  \State $X \leftarrow \texttt{work}(Y)$: The miner set $Y$ finds FPoW $X$
  \State $Y \leftarrow \texttt{submit}(X)$: FPoW $X$ is submitted to the manager of $Y$
  \State $\texttt{publish}(Y, X)$: The manager of $Y$ publishes FPoW $X$
  \State $\texttt{discard}(X)$: An attacker discards FPoW $X$
  \medskip
  \Function{round}{}
    \State $k=1$
    \Foreach{$P_i \subset P$}
      \State $F_{wh, i}=\emptyset$
      \BState \emph{Generate a Fork}:
      \If {$F_k \leftarrow \texttt{work}(A\cap P^c)$}
        \State $\texttt{publish}(A,F_k)$ \Comment{Case A}
      \ElsIf {$F_k \leftarrow \texttt{work}(A^c\cap P_i)$}
        \State $P_i \leftarrow \texttt{submit}(F_k)$
        \State $\texttt{publish}(P_i,F_k)$ \Comment{Case B}
      \ElsIf {$F_k \leftarrow \texttt{work}(A^c\cap P^c)$}
        \If {$F_{wh, i}\ne\emptyset$}
          \State $\texttt{publish}(A^c\cap P^c,F_k)$
          \State $P_i \leftarrow \texttt{submit}(F_{wh, i})$
          \State $\texttt{publish}(P_i,F_{wh, i})$ \Comment{Fork, Case C, D}
        \Else 
          \State $\texttt{publish}(A^c\cap P^c,F_k)$ \Comment{Case E}
        \EndIf
      \Else 
          \State $F_k \leftarrow \texttt{work}(A\cap P_i)$
	  \If {$F_{wh, i}=\emptyset$}
            \State $F_{wh, i}=F_k$
	  \Else
	    \State $\texttt{discard}(F_k)$
	\EndIf
        \State $k$++
        \State goto \emph{Generate a Fork}
      \EndIf
    \EndForeach
  \EndFunction
\end{algorithmic}
\end{algorithm}
\if false
\vspace{15pt}

\captionof{algorithm}{FAW attack game among two pools\label{FAWgame}}
\begin{algorithmic}[1]
  \State $A$: An attacker
  \Function{FAW$_{multiple}$}{$n$}
      \State $P_j$: a\;target\;pool $j$
    \State $P: \bigcup P_j$
    \State $F_k$: The $k$-th FPoW\;found\;in\;one\;round
    \State $F_{hold, i}$: The FPoW found by $A$ in pool $i$
    \While {true}
      \State $A$ starts the FAW attack against $P$ for $i$-th round
      \If {$A$ finds $F_1$ by honest mining}
        \State Propagate $F_1$
      \ElsIf {$A^c$ finds $F_1$}
        \State Continue
      \Else \Comment{$A$ finds $F_1$ in $P$}
        \State {Add $F_1$ to $F_{list}$}
        \BState \emph{Generate a Fork}:
        \State Hold $F_{list}$ and continue mining
        \If {$(A \bigcup P)^c$ finds new $F_k$}
          \State {Submit all $F_{list}$ to each pool $P_j$}
          \ElsIf {$(P-A)$ finds new $F_k$}
            \State {Discard $F_{list}$}
          \ElsIf {$A$ finds new $F_k$ by honest mining}
            \State Propagate $F_k$ and discard $F_{list}$
          \Else \Comment{$A$ finds new $F_k$ in $P_j$}
          \If {$A$ already found $F_j$ in $P_j$}
            \State Discard $F_k$
          \Else \Comment{$F_{list}$ is not found in $P_j$}
            \State Add $F_k$ to $F_{list}$
          \EndIf
          \State goto \emph{Generate a Fork}
        \EndIf
      \EndIf
      \State $i=i+1$
    \EndWhile
  \EndFunction
\end{algorithmic}
\fi

\end{appendices}

\end{document}